\documentclass[fleqn,usenatbib]{mnras}

\usepackage{newtxtext,newtxmath}
\usepackage[T1]{fontenc}
\usepackage{ae,aecompl}
\usepackage{times}

\usepackage{graphicx}	% Including figure files
\usepackage{amsmath}	% Advanced maths commands
\newcommand{\Mpc}{\ensuremath{~\mathrm{Mpc}}}
\newcommand{\hMpc}{\ensuremath{~h^{-1}\mathrm{Mpc}}}

\renewcommand{\deg}{\ensuremath{~\mathrm{deg}}}
\renewcommand{\arcmin}{\ensuremath{~\mathrm{arcmin}}}
\newcommand{\degSq}{\ensuremath{~\mathrm{deg}^2}}
\newcommand{\map}[1]{\ensuremath{#1 \times #1 \degSq}}

\newcommand{\cosmoslics}{\emph{cosmo}-SLICS}
\newcommand{\LSST}{\textsc{lsst}}

\title[Cosmology with WL peaks]{Cosmological forecasts with the clustering of weak lensing peaks}

\author[Davies et. al] 
{Christopher T. Davies,$^{1}$\thanks{E-mail:christopher.t.davies@durham.ac.uk (CTD)}
Marius Cautun,$^{2}$
Benjamin Giblin,$^{3}$
Baojiu Li,$^{4}$\newauthor
Joachim Harnois-Déraps,$^{5,6}$
and Yan-Chuan Cai$^{3}$
\\
$^{1}$Faculty of Physics, Ludwig-Maximilians-Universität, Scheinerstr. 1, 81679 Munich, Germany\\
$^{2}$Leiden Observatory, Leiden University, PO Box 9513, NL-2300 RA Leiden, the Netherlands\\
$^{3}$Scottish Universities Physics Alliance, Institute for Astronomy, University of Edinburgh, Blackford Hill, Scotland, UK\\
$^{4}$Institute for Computational Cosmology, Department of Physics, Durham University, South Road, Durham DH1 3LE, UK\\
$^5$School of Mathematics, Statistics and Physics, Newcastle University, Herschel Building, NE1 7RU, Newcastle-upon-Tyne, UK\\
$^6$Astrophysics Research Institute, Liverpool John Moores University, 146 Brownlow Hill, Liverpool, L3 5RF, UK\\
}

\date{Accepted XXX. Received YYY; in original form ZZZ}

\pubyear{2021}

\begin{document}
\label{firstpage}
\pagerange{\pageref{firstpage}--\pageref{lastpage}}
\maketitle

\begin{abstract}
Maximising the information that can be extracted from weak lensing measurements is a key goal for upcoming stage IV surveys. This is typically achieved through statistics that are complementary to the cosmic shear two-point correlation function, the most well established of which is the weak lensing peak abundance. 
In this work, we study the clustering of weak lensing peaks, and present parameter constraint forecasts for an \LSST{}-like survey. We use the \cosmoslics{} $w$CDM simulations to measure the peak two-point correlation function for a range of cosmological parameters, and use the simulation data to train a Gaussian process regression emulator which is applied to generate likelihood contours and provide parameter constraint forecasts from mock observations. 
We investigate the dependence of the peak two-point correlation function on the peak height, and find that the clustering of low amplitude peaks is complementary to that of high amplitude peaks. Consequently, their combination gives significantly tighter constraints than the clustering of high peaks alone. 
The peak two-point correlation function is significantly more sensitive to the cosmological parameters $h$ and $w_0$ than the peak abundance, and when the probes are combined, constraints on $\Omega_{\rm m}$, $S_8$, $h$ and $w_0$ improve by at least a factor of two, relative to the peak abundance alone.
Finally, we compare the forecasts for weak lensing peaks and weak lensing voids, and show that the two are also complementary; both probes can offer better constraints on $S_8$ and $w_0$ than the shear correlation function by roughly a factor of two.

\end{abstract}


%%%%%%%%%%%%%%%%%%%%%%%%%%%%%%%%%%%%%%%%%%%%%%%%%%

%%%%%%%%%%%%%%%%% BODY OF PAPER %%%%%%%%%%%%%%%%%%

\section{Introduction}

The standard cosmological model, $\Lambda{\rm CDM}$, consists of matter dominated by cold dark matter (CDM), and a late-time accelerated expansion that is driven by a positive cosmological constant $\Lambda$. This model is highly successful at describing a number of independent observations, which constrain the $\Lambda{\rm CDM}$ parameters with a large degree of concordance. Notably, measurements of fluctuations in the Cosmic Microwave Background (CMB) \citep{Planck2018} have been used to constrain cosmological parameters including the present-day expansion rate of the Universe, $H_{0}$, the matter density parameter, $\Omega_{\rm m}$, and the matter fluctuation amplitude $\sigma_8$, defined as the root-mean-squared matter density perturbations smoothed on $8h^{-1}\Mpc$ scales. 

%\JHD{[I believe we should use $\Omega_{\rm m}$, not $\Omega_{m}$ in MNRAS.]}

Gravitational lensing is another promising observational probe that can be used to constrain many cosmological parameters, where the light of distant source images is distorted by the gravitational potentials of the foreground matter. In the weak lensing (WL) regime, light is deflected by the large scale structure (LSS) of the Universe, and the weak lensing signal is measured through the correlations in distortions of many source galaxies \citep{Bacon2000,Kaiser2000,VanWaerbeke2000,Wittman2000}. This allows us to probe the total matter distribution of the Universe on the largest scales \citep[see][for reviews]{Bartelmann2001,Kilbinger2015}, and offers a powerful method to study the properties of dark matter and dark energy.

Recent WL observations that supplement the CMB parameter measurements include the Dark Energy Survey (DES) \citep{DES2021} \footnote{\url{https://www.darkenergysurvey.org/}}, Hyper Supreme-Cam (HSC) \citep{Hikage2019} \footnote{\url{https://hsc.mtk.nao.ac.jp/ssp/}} and the Kilo-Degree Survey (KiDS) \citep{Asgari2020} \footnote{\url{http://kids.strw.leidenuniv.nl/}} WL surveys. All of these surveys measure lower values of $\sigma_8$ compared to Planck, with a statistically significant disagreement arising between the Planck and KiDS constraints. However, the more recent results from DES are statistically compatible with Planck. This is one example where different observations point to slightly different values of certain cosmological parameters, which, in more extreme cases, may imply the presence of either unaccounted for systematics or new physics which is not modelled. An example of a larger discrepancy, leading to a parameter tension, is the $H_0$ tension, where multiple observations find that measurements from the early Universe are broadly inconsistent with those of the late Universe \citep{Verde2019}, particularly the distance scale measurement of $H_0$ based on Cepheids by the SH0ES collaboration \citep{Riess:2019cxk}. 

In order to address these parameter tensions, and more deeply probe the nature of the Universe, it is important to measure cosmological parameters as precisely as possible. This can be achieved by maximising the information that can be extracted from a given survey. The standard approach for weak lensing surveys is to measure $\Lambda\rm{CDM}$ parameters with two-point statistics, such as the shear-shear correlation function  \citep{Schneider2002,Semboloni2006,Hoekstra2006,Fu2008,Heymans2012,Kilbinger2013,Hildebrandt2017,Troxel2018,Hikage2019,Aihara2019,Asgari2020,DES2021}. However, the shear two-point correlation function (2PCF) does not capture non-Gaussian information, and, due to the non-linear evolution of the Universe, weak lensing data are highly non-Gaussian. To fully exploit the data, many non-Gaussian statistics have been developed, which encapsulate information beyond two-point statistics. A well-established example is the abundance of WL peaks (local maxima in the convergence field), which has been shown to be complementary to the shear two-point function and helps break the $\Omega_{\rm m}$-$\sigma_8$ parameter degeneracy \citep{Jain2000,Pen2003,Dietrich2010}. Peaks are also shown to outperform the standard methods for constraining the sum of neutrino mass \citep{Li2018} and $w_0$ \citep{Martinet2020}, and can be used to provide constraints on modified gravity theories \citep{X.Liu2016}. When used in conjunction with the shear two point correlation function, WL peaks have been used to provide the tightest constraints on $S_8$ from DES-Y1 WL data \citep{Harnois2020}. WL peaks also offer utility for other non-Gaussian statistics, such as WL voids \citep{Davies2018,Davies2020b}, where the peaks can be used as tracers to identify the voids. By including WL peaks as a complementary statistic, the measurement errors on cosmological parameters can be reduced, which will help inform the statistical significance of any parameter tensions between multiple observations. 

When used to constrain cosmological parameters, peak analyses typically %only 
focus on the WL peak abundance, % function, 
which is the number density of WL peaks as a function of their lensing
amplitude. Studies have shown that the WL peaks with the highest amplitudes tend to correspond to large haloes along the line of sight \citep{Hamana2004,J.Liu2016,Wei2018}. For this reason, WL peaks identified in surveys such as HSC can be used to search for galaxy clusters \citep[e.g.,][]{Hamana2020}. Furthermore, shear 2PCF measurements are typically combined with measurements of galaxy clustering (and galaxy-galaxy lensing) to further tighten the cosmological constraints \citep[e.g.,][]{DES2021}. So, if WL peaks correspond to massive haloes, and hence massive galaxies, and the clustering of these galaxies is known to contain complementary information, then studying and exploiting the clustering of WL peaks is a natural next step in maximising the utility of WL peaks. 

Previously, \cite{Marian2013} have shown that the 2PCF of WL peaks with high lensing amplitudes does not contain much complementary information to the peak abundance alone. In \cite{Davies2019}, we presented some simple scaling relations for the WL peak 2PCF, and found that the clustering of low-amplitude WL peaks also appears to be sensitive to the cosmological parameters $\Omega_{\rm m}$ and $\sigma_8$. In this work, with a more detailed analysis, we show that the clustering of low-amplitude peaks contains significant complementary information to the clustering of high peaks, and that when the clustering of multiple peak height ranges are combined, the WL peak 2PCF offers similar constraining power to the peak abundance alone, where the two probes have different degeneracy directions. Therefore, we also show that, when their abundance and 2PCF are combined, the total cosmological information that is extracted from the WL peaks is significantly improved.

We use the numerical simulation suite, \cosmoslics{} \citep{Harnois2019} to measure the peak abundance and 2PCF, for a range of cosmological parameters. This data is then used to train a Gaussian process regression emulator, which, combined with Markov chain Monte Carlo, allows us to generate likelihood contours and provide forecast parameter constraints for an \LSST-like survey.

The layout of the paper is as follows. In Section \ref{sec:theory} we outline the relevant theory for our WL peak analysis. In Section \ref{sec:Mock data etc} we describe how we generate our mock observational data, and our emulation and likelihood analysis pipeline. In Section \ref{sec:weak lensing peak statistics} we present the WL peak statistics used in the analysis, first from the mock data, and then from the emulator, in order to understand how these statistics depend on the cosmological parameters $\Omega_{\rm m}$, $S_8$, $h$ and $w_0$. In Section \ref{sec:param constraints forecast} we present the parameter constraint forecasts for the WL peak 2PCF and peak abundance. Finally, our conclusions are presented in Section \ref{sec:discussion and conclusions}. We also have two appendices where we present the covariance matrix used in our analysis, and study the accuracy of our emulator. 

\section{Theory}\label{sec:theory}

The lens equation for a gravitationally-lensed image, relating the deflection angle $\pmb{\alpha}$ to the true position of the source $\pmb{\beta}$ and the observed position of the image $\pmb{\theta}$, is
\begin{equation}
    \pmb{\alpha} = \pmb{\beta} - \pmb{\theta} \, .
\end{equation}
Neglecting second-order effects, the deflection angle is the gradient of a 2D lensing potential $\psi$,
\begin{equation}
    \pmb{\alpha} = \pmb{\nabla}\psi \, ,
    \label{eq:alpha}
\end{equation}
where $\psi$ is given by
\begin{equation}
    \psi(\pmb{\theta},\chi) = \frac{2}{c^2} \int_0^{\chi} \frac{f_K(\chi - \chi')}{f_K(\chi) f_K(\chi')} \Phi(\chi' \pmb{\theta},\pmb{\theta}) d\chi' \, .
    \label{eq:lensing potential}
\end{equation}
Here, $\chi$ is the comoving distance from the observer to the source and $\chi'$ is the comoving distance from the observer to the continuously-distributed lenses. $f_K(\chi)$ is the comoving angular distance, with the spatial curvature of the universe denoted by $K$. Note that for a flat universe with $K=0$ (as used in this work) $f_K(\chi) = \chi$. $\Phi$ is the 3D lensing potential of the lens, and $c$ the speed of light. $\Phi$ is given by the Poisson equation 
\begin{equation}
    \nabla^2\Phi = 4 \pi G a^2 \bar{\rho} \delta \, ,
    \label{eq:Poisson equation}
\end{equation}
where $\rho$ is the matter density of the Universe (with the mean denoted by a bar), $\delta \equiv \rho/\bar{\rho} - 1$, $a$ is the scale factor and $G$ is the gravitational constant. 

%Eq.~\eqref{eq:lensing potential} shows that the lensing potential is a line-of-sight integral of the matter distribution from the source to the observer. The contribution that matter at distance $\chi'$ along the line of sight makes to the total lensing potential is weighted by $(\chi - \chi') / \chi \chi'$ and so depends on its distances from the source and observer.

%Eq.\eqref{eq:alpha} allows Eq. \eqref{eq:amp mat} to be expressed in terms of $\psi$
%\begin{equation}
%    A_{ij} = \delta_{ij} - \partial_{i} \partial_{j} \psi \, ,
%\end{equation}
%where partial derivatives are taken with respect to $\pmb{\theta}$. The  matrix $\pmb{A}$ can be parameterised through the more physically instructive terms convergence, $\kappa$, and shear, $\gamma = \gamma_1 + i\gamma_2$, as
%\begin{equation}
%    \pmb{A} = 
%%    \begin{pmatrix}
 %   1 - \kappa -\gamma_1 & -\gamma_2\\
 %   -\gamma_2 & 1-\kappa+\gamma_1
 %   \end{pmatrix}
 %   \, .
%\end{equation}    
The convergence $\kappa$ and shear $\gamma = \gamma_1 + i\gamma_2$ can be related to the lensing potential via
\begin{equation}
    \kappa \equiv \frac{1}{2} \nabla^2_{\pmb{\theta}} \psi \, ,
    \label{eq:convergence}
\end{equation}
and
\begin{equation}
    \gamma_1 \equiv \frac{1}{2}\left(\nabla_{\pmb{\theta}_1}\nabla_{\pmb{\theta}_1}-\nabla_{\pmb{\theta}_2}\nabla_{\pmb{\theta}_2}\right)\psi, 
    \quad\quad\quad
    \gamma_2 \equiv \nabla_{\pmb{\theta}_1}\nabla_{\pmb{\theta}_2}\psi,
    \label{eq:shear}
\end{equation}
where $\nabla_{\pmb{\theta}} \equiv (\chi)^{-1}\nabla$. Eqs.~\eqref{eq:lensing potential}, \eqref{eq:Poisson equation} and ~\eqref{eq:convergence} allows us to express the convergence in terms of the matter overdensity
\begin{equation}
    \kappa(\pmb{\theta},\chi) = \frac{3H_0^2\Omega_{\rm{m}}}{2c^2}\int_0^{\chi}\frac{f_K(\chi - \chi')}{f_K(\chi)} f_K(\chi') \frac{\delta(\chi'\pmb{\theta},\chi')}{a(\chi')} d\chi' \, .
    \label{eq:conv source}
\end{equation}
%This shows that the observed WL convergence can be interpreted as the projected density perturbation along the line of sight, weighted by the lensing efficiency factor $(\chi-\chi')\chi'/\chi$. Here, the lensing efficiency is greatest at $\chi' = \chi / 2$, when the lens is halfway between the source and the observer. 

The above derivation uses a fixed source plane. However, in real WL observations, the catalogue of source galaxies has a probability distribution $n(\chi)$ that spans over a range of $\chi$ values, and Eq. \eqref{eq:conv source} is then weighted by this distribution to obtain $\kappa(\pmb{\theta})$ \citep[see, e.g.,][for details]{Kilbinger2015}
\begin{equation}
    \kappa(\pmb{\theta}) = \int_0^{\chi} n(\chi') \kappa(\pmb{\theta},\chi') d\chi' \, .
\end{equation}

WL observations rely on accurately measuring the shapes of galaxies, and cross correlating the shapes of neighbouring galaxies. However, measurements of the galaxy shapes used to extract the lensing signal are dominated by the random galaxy shapes and orientations, which is referred to as galaxy shape noise (GSN).  
Since the lensing signal is weak, when identifying WL peaks (local maxima in the convergence field $\kappa(\pmb{\theta})$) it is convenient to express the convergence relative to the standard deviation of the corresponding GSN component of the field, $\sigma_{\rm{GSN}}$. This is given by the following definition of signal-to-noise,
\begin{equation}
    \nu = \frac{\kappa}{\sigma_{\rm{GSN}}}\, ,
    \label{eq:nu}
\end{equation}
where $\sigma_{\rm{GSN}}$ is the standard deviation of the contributions to the signal from galaxy shape noise. The $\sigma_{\rm{GSN}}$ term can be calculated by generating mock GSN maps using the prescription from \cite{Jain2000} and \cite{VanWaerbeke2000} and applying to them any transformations also applied to the convergence maps, such as smoothing. Mock GSN maps are generated by assigning to pixels random convergence values from a Gaussian distribution with standard deviation 
\begin{equation}
        \sigma_{\rm{pix}}^2 = \frac{\sigma_{\rm{int}}^2}{2 \theta_{\rm{pix}}^2 n_{\rm{gal}} }
    \;,
    \label{eq: GSN gaussian}
\end{equation}
where $\theta_{\rm{pix}}$ is the width of each pixel, $\sigma_{\rm{int}}$ is the intrinsic ellipticity dispersion of the source galaxies, and $n_{\rm{gal}}$ is the measured source galaxy number density. In this work we use $\sigma_{\rm{int}} = 0.28$ and $n_{\rm{gal}} = 20 $ arcmin$^{-2}$ as will be discussed in Section \ref{sec:Numerical simulations}.

%It will also be useful to compare constraints from Eq. \ref{eq:gamma_t} to the standard shear two-point correlation function, which is given by

%\begin{equation}
%    \xi_{\pm}(\pmb{\theta}) = \langle \gamma_t \gamma_t \rangle \pm \langle \gamma_{\times} \gamma_{\times} \rangle = \frac{1}{2\pi} \int_0^{\infty} dl l P_{\kappa}(l) J_{0,4}(l\pmb{\theta}) \, ,
%\end{equation}

%where $\gamma_t = -\mathbb{R}(\gamma e^{-2i\phi})$ (equivalent to Eq. \eqref{eq:gamma_t}, but presented for completeness), $\gamma_{\times} = -\mathbb{I}(\gamma e^{-2i\phi})$, $\phi$ is the polar angle of the separation vector $\pmb{\theta}$ , $J_0$ and $J_4$ are the Bessel functions for $\xi_{+}$ and $\xi_{-}$ respectively, and $l$ is the Fourier mode.

As mentioned in the introduction, WL peaks are closely related to the dark matter haloes along the line of sight. In cosmology, both the abundance and large-scale clustering of haloes encode useful information about the underlying cosmological model and parameter values. Therefore, as well as studying the abundance of WL peaks, we will also study their clustering. The extent to which objects are clustered can be measured through the two-point correlation function (2PCF) which is defined as the excess probability, relative to a random distribution, of finding a pair of objects at a given separation $\theta$. Formally, this is written as
\begin{equation}
    dP_{ij}(\theta) = \overline{n}^2(1+\xi(\theta))dA_i dA_j \, ,
\end{equation}
where $\overline{n}$ is the expected tracer number density, $dA_i$ and $dA_j$ are two sky area elements that are separated by a displacement $\boldsymbol{\theta}$ with amplitude $\theta$, and $\xi(\theta)$ is the 2PCF. We have $\xi(\boldsymbol{\theta}) = \xi(\theta)$ thanks to statistical isotropy. In practice, the 2PCF can be measured through the Landy-Szalay estimator \citep{Landy1993} which requires the generation of matching catalogues containing randomly distributed points and is given by
\begin{equation}
    \xi_{\rm{LS}}(\theta) = 1 + \bigg(\frac{N_R}{N_D}\bigg)^2 \frac{DD(\theta)}{RR(\theta)} - \bigg(\frac{N_R}{N_D}\bigg) \frac{DR(\theta)}{RR(\theta)}.
    \label{Eq: LS estimator}
\end{equation}
In the above $N_D$ and $N_R$ are the numbers of data and random points, and $DD$, $DR$ and $RR$ are the numbers of data-data, data-random and random-random pairs in bins $\theta \pm \delta \theta$, respectively. See \citet{Davies2019} for more details about the measurement of the peak 2PCF, which are important for small lensing maps.

\section{Methodology}
\label{sec:Mock data etc}

In this section we describe the methodology followed in this work, including the simulations, mock lensing data, emulation and likelihood analysis.

\subsection{Mock Data}
\label{sec:Numerical simulations}

In this work we use the SLICS and \cosmoslics{} \citep{Harnois2015,Harnois2018,Harnois2019} mock WL convergence maps, which we briefly outline in this subsection.

The \cosmoslics{} are a suite of high-resolution $N$-body simulations that were run for 26 sets of $[\Omega_{\rm m},S_8,h,w_0]$ cosmological parameters. Here $S_8 \equiv \sigma_8 (\Omega_{\rm m}/0.3)^{0.5}$, $h = H_0 / 100 {\rm kms}^{-1} \Mpc^{-1}$ is the reduced Hubble constant, and $w_0$ the dark energy equation-of-state parameter, which is assumed to be a constant that is allowed to deviate from $-1$ (cosmological constant).

The four dimensional parameter space is sampled using a Latin hypercube, which samples the parameter space comprehensively with a low node count. The exact cosmological parameter space that is probed by the \cosmoslics{} is shown in Fig.~\ref{fig:cosmoSLICS nodes}. Each simulation volume is a cube with length $L = 505 \hMpc$, with $N = 1536^{3}$ dark matter particles. To reduce the impact of cosmic variance, two simulations are run for each cosmology, starting from different (paired) initial conditions. For each set of cosmological parameters, 50 pseudo-independent light-cones are constructed by resampling projected mass sheets, which are then ray-traced under the Born approximation to construct lensing maps and catalogues \citep[see][for full details about the light-cone and catalogue construction]{Harnois2019}. 

We use the {\it Generic} \cosmoslics{} source catalogue selected over the range $z_s = [0.6,1.4]$  to match  $\LSST$ specifications, which gives a conservative source galaxy number density of $20 \arcmin^{-2}$. From this we generate 50 WL convergence maps for each of the nodes, with a sky coverage of \map{10} each and $3600^{2}$ pixels, following the method described in  \citet{Giblin2018}. These maps are then smoothed with a Gaussian filter with smoothing scale $\theta_s = 1 \arcmin$. \citet{Davies2019} contains a study of the impact of different smoothing scales on the WL peak 2PCF.

For estimates of the covariance matrices, we use the SLICS suite to produce 615 WL convergence maps at the fiducial cosmology, which match the properties of the \cosmoslics{} maps. These are obtained from fully independent $N$-body realisations carried out at the same cosmology\footnote{The SLICS cosmology has the following parameter values: [$\Omega_{\rm m}$, $\sigma_8$, $h$, $w_0$, $n_{\rm s}$, $\Omega_{\rm b}$] = [0.2905, 0.826, 0.6898, -1.0, 0.969, 0.0474].}, but with different seeds in their initial conditions, allowing to accurately capture the sample variance. The larger number of SLICS realisations relative to \cosmoslics{} allows us to calculate robust covariance matrices and to use large data vectors in the likelihood analysis below when combining probes. 

\subsection{Emulation and likelihood analysis}
\label{sec:Emulation}

In this subsection, we outline the procedure used to test the sensitivity of WL peak statistics to the cosmological parameters $\Omega_{\rm m}$, $S_8$, $h$ and $w_0$. 

\begin{figure}
    \centering
    \includegraphics[width=\columnwidth]{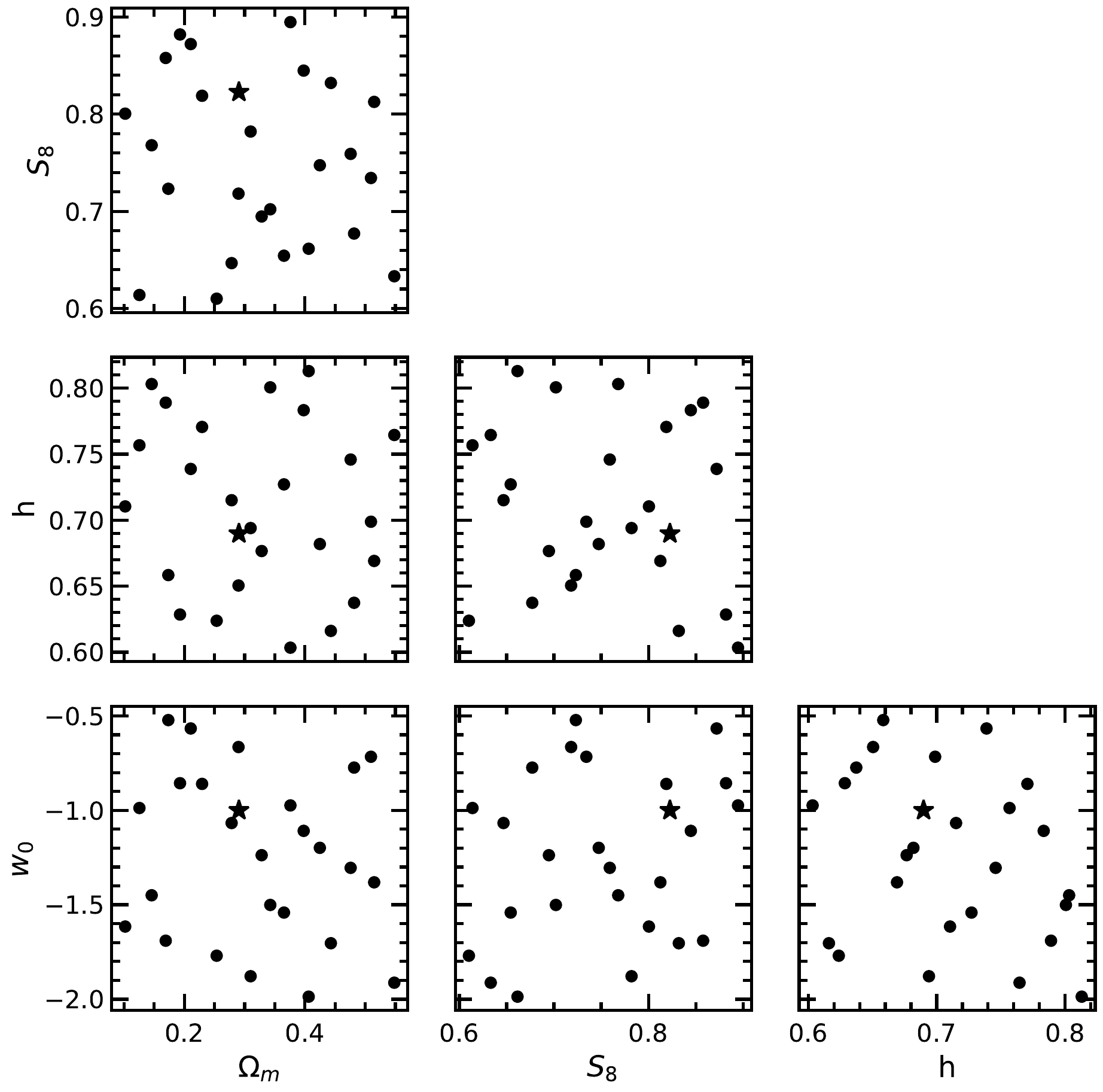}
    \caption{The 4-dimensional parameter space ($[\Omega_{\rm m},S_8,h,w_0]$) sampled by the \cosmoslics{} simulation suite. The fiducial cosmology is indicated by a star with parameter values [$0.29$, $0.82$, $0.69$, $-1.00$].}
    \label{fig:cosmoSLICS nodes}
\end{figure}

First, we measure the WL peak statistics from the 50 convergence maps for each of the nodes shown in Fig.~\ref{fig:cosmoSLICS nodes}. Then, in order to predict the WL peak statistics at arbitrary points in this parameter space, we use the Gaussian process (GP) regression emulator from \textsc{scikit}-\textsc{learn} \citep{Pedregosa2011} to interpolate the peak statistics between nodes. GP regression is a non-parametric Bayesian machine learning algorithm used to make probabilistic predictions that are consistent with the training data \citep[see, e.g.,][for some of its early applications in cosmology]{Habib2007,Schneider2008}. The accuracy of the GP emulator trained on \cosmoslics{} has been tested extensively and shown to reach a few percent level in predictions of weak lensing shear two-point correlation functions \citep{Harnois2019}, density split statistics \citep{Burger2020}, persistent homology statistics \citep{Heydenreich2020}, aperture mass statistics \citep{Martinet2020} and WL void statistics \citep{Davies2020b}. In this work the average peak statistics and their standard errors at each node are used as the training data. We present results of the accuracy of the emulator for the peak statistics in Appendix \ref{app:accuracy}. 

Finally, once the emulator has been trained and tested, we use Monte Carlo Markov Chain (MCMC) to estimate the posteriors of the parameters for the entire parameter space and produce likelihood contours. We use the \textsc{emcee} Python package \citep{Foreman2013} to conduct the MCMC analysis in this work sampling the 4D parameter space as follows. We employ a Bayesian formalism, in which the likelihood of the set of cosmological parameters $\pmb{p} = [\Omega_{\rm m},S_8,h,w_0]$ given a data set $\pmb{d}$, is given by
\begin{equation}
    P(\pmb{p}|\pmb{d}) = \frac{P(\pmb{p}) P(\pmb{d}|\pmb{p})}{P(\pmb{d})} \, ,
\end{equation}
where $P(\pmb{p})$ is the prior, $P(\pmb{d}|\pmb{p})$ is the likelihood of the data conditional on the parameters, and $P(\pmb{d})$ is the normalisation. In this work we use flat priors with the following upper and lower limits respectively for $\Omega_{\rm m}$: [0.10, 0.55], $S_8$: [0.61, 0.89], $h$: [0.60, 0.81],
$w_0$: [-1.99, -0.52], which matches the parameter space sampled by \cosmoslics{}. 
%\JHD{[We should say here that we are using a multivariate Gaussian likelihood. This follows previous data analyses and should be fine, though not perfect. Not sure how deep we want to push this discussion here]}
The log likelihood can be expressed as
\begin{equation}
    \log(P(\pmb{d}|\pmb{p})) = -\frac{1}{2} \left[\pmb{d} - \mu(\pmb{p})\right]C^{-1}\left[\pmb{d} - \mu(\pmb{p})\right] \, ,
    \label{eq:log likelihood}
\end{equation}
where $\mu(\pmb{p})$ is the prediction generated by the emulator for a set of parameters $\pmb{p}$, and $C^{-1}$ is the inverse of the covariance matrix. We use the emulator's prediction of a statistic at the fiducial cosmology as the data $\pmb{d}$. This choice is for simplicity and presentation purposes, which ensures that the confidence intervals are always centred on the true values of the cosmological parameters allowing for easier comparisons between multiple probes. 

The likelihood returns a 4D probability distribution that indicates how well different regions of the parameter space match the input data $\pmb{d}$. Note that Eq.~\eqref{eq:log likelihood} assumes that the covariance matrix does not depend on the cosmological parameters. 

We use the 615 SLICS WL map realisations (which match the fiducial cosmology) to calculate the covariance matrices, and then divide it by a factor of $180$ to rescale the covariance matrix from a $100$ deg$^{2}$ area to the \LSST{} survey area, which we take as $18,000$ deg$^{2}$. The joint covariance matrix for the peak probes studied in this work is presented in Appendix \ref{app:cov}. We also multiply the inverse covariance matrix by a debiasing factor $\alpha$, which accounts for the bias introduced when inverting a noisy covariance matrix \citep{Anderson2003,Hartlap2007}, and is given by:
\begin{equation}
    \alpha = \frac{N - N_{\rm{bin}} - 2}{N - 1} \, .
\end{equation}
Here $N = 615$ is the number of WL maps used to calculate the covariance matrix and $N_{\rm{bin}}$ is the number of bins used to measure the statistic. %We note however that \cite{Sellentin2016} present an alternative approach to robustly account for the uncertainty in the estimated covariance, via a student-$t$ likelihood distribution.

\section{Weak lensing peak statistics}
\label{sec:weak lensing peak statistics}

In this section, we present the weak lensing peak statistics studied in this work, which include the peak abundance and the peak 2PCF. For each statistic, we first show their measurements from the \cosmoslics{} nodes which are used as the training data for the emulator. We then present emulations of the statistic by varying one cosmological parameter at a time, to exemplify its sensitivity to different cosmological parameters, which will aid the interpretation of the forecast cosmological constraints in Section \ref{sec:param constraints forecast}.

\subsection{Weak lensing peak abundance}\label{subsec:peak abubdance}
\begin{figure}
    \centering
    \includegraphics[width=\columnwidth]{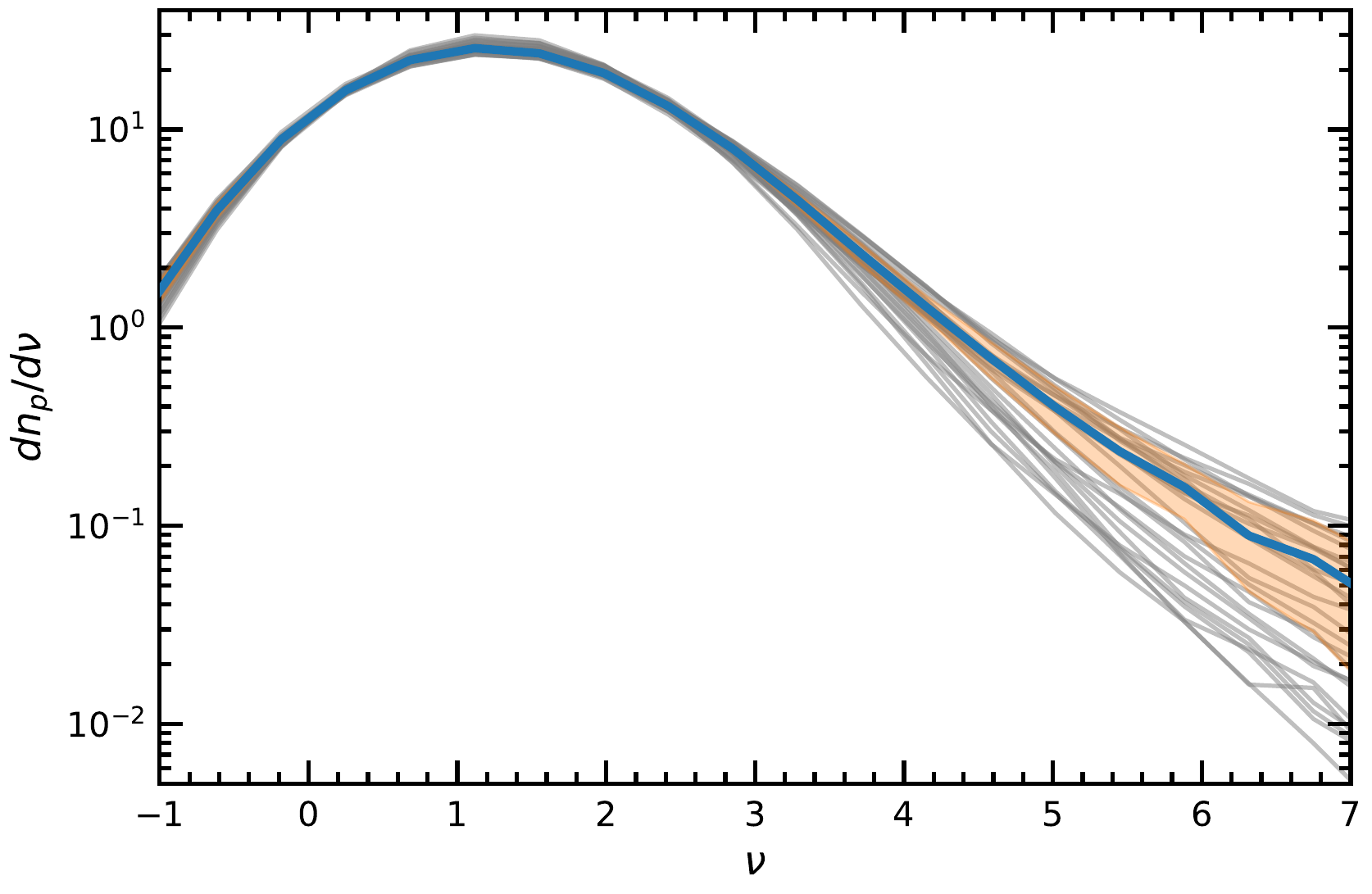}
    \caption{(Colour Online) The differential WL peak number density (abundance) as a function of peak height $\nu$. The grey curves correspond to the 26 \cosmoslics{} nodes in Fig.~\ref{fig:cosmoSLICS nodes}, with the fiducial cosmology plotted as a blue thick curve. The orange shaded regions shows the 1$\sigma$ standard error measured from the 50 fiducial cosmoSLICS realisations, multiplied by a factor of ten to increase visibility.}
    \label{fig:PA}
\end{figure}

\begin{figure*}
    \centering
    \includegraphics[width=2\columnwidth]{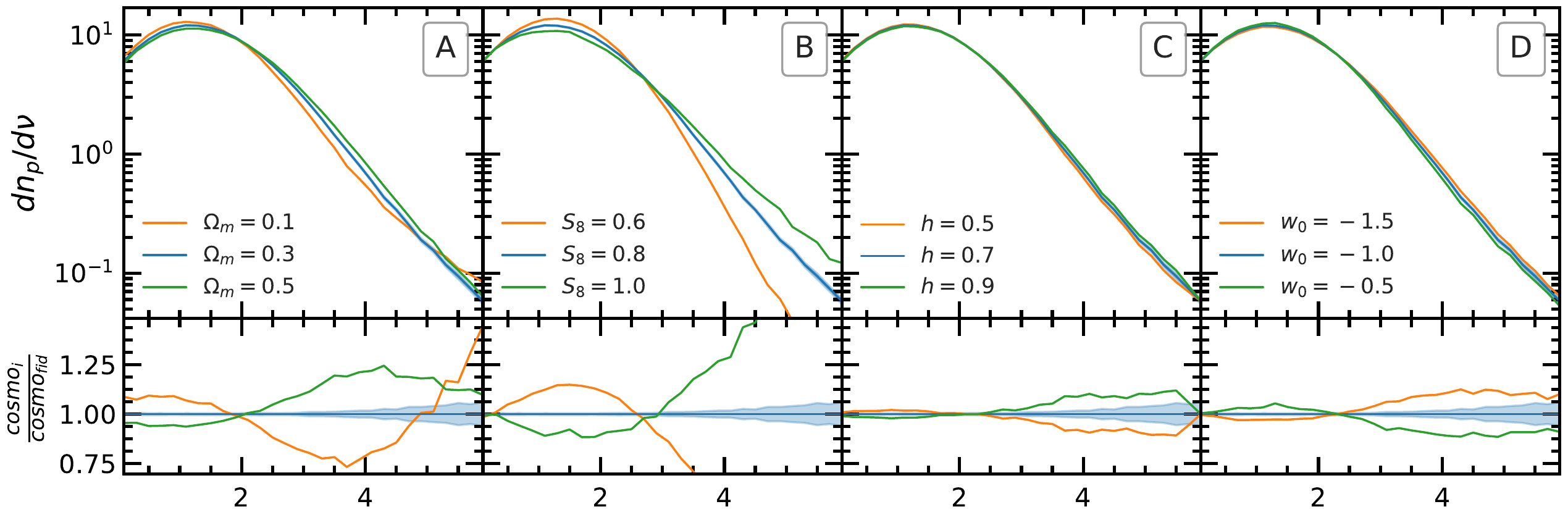}
    \caption{(Colour Online) \textit{Top row}: the emulated peak abundance. The curves correspond to the cosmological parameters [$\Omega_{\rm m}$, $S_8$, $h$, $w_0$] with values [$0.3$, $0.8$, $0.7$, $-1.0$], unless otherwise stated in the sub-panel legends. Each sub-panel corresponds to varying one cosmological parameter at a time, denoted in the legend.
    \textit{Bottom row}: The curves from the top row, divided by the fiducial cosmology (blue curve in the top row).
    The $1\sigma$ standard errors measured from the $50$ \cosmoslics{} realisations are included on the fiducial (blue) curves, and they are barely visible in the upper sub-panels.
    }
    \label{fig:PA vary 1 param}
\end{figure*}

Fig.~\ref{fig:PA} shows the differential WL peak abundance (number density) measured in each of the 26 nodes in Fig.~\ref{fig:cosmoSLICS nodes}. The abundance of the fiducial cosmology is shown by the blue curve, with the rest of the cosmologies plotted in grey.

First, the figure shows that there are an appreciable number of peaks with amplitudes below $\nu$ = 0, which correspond to local maxima in regions that are underdense. In this analysis we do not use peaks with $\nu < 0$ for our forecast constraints for the following two reasons. First, \cite{Martinet2018} have shown that peaks with $\nu<0$ correlate very strongly with peaks with $\nu>0$, and so there is little gain in parameter constraints when these peaks are included. Second, due to their low amplitude, these peaks are also more likely to be affected by GSN. 

The maxima of the peak abundance occurs just above $\nu = 1$ for all cosmologies. Due to the low signal-to-noise ratio, this indicates that a large fraction of the total number of peaks correspond to spurious local maxima induced by galaxy shape noise, rather than a physical signal induced by matter overdensities along the line of sight, while more peaks at the high-$\nu$ end are produced by a true physical signal.

As the peak abundance approaches higher $\nu$ values, the spread in the peak abundances between different cosmologies increases significantly, where the abundance of peaks at $\nu = 6$ can differ by an order of magnitude between the most extreme cosmologies. This is due to two factors. First, because the peaks at this amplitude are not dominated by noise, differences in the physical signal are more visible. Second, because the high-mass end of the halo mass function varies more significantly as a function of cosmological parameters, so does the peak abundance, since the largest peaks are created by the largest haloes \citep{J.Liu2016,Wei2018}.

Whilst the high-$\nu$ end of the abundance exhibits the greatest variation amongst the different cosmological parameters, this region also has the highest sample variance, since high peaks are orders of magnitude less abundant than low peaks. Therefore, as $\nu$ increases, the increased spread between cosmologies is in direct competition with the increased statistical uncertainty. For this reason it is important to consider the abundance of peaks over a wide $\nu$ range in our forecasts. We also note that large $\nu$ peaks are more affected by uncertainties such as intrinsic alignments and baryonic physics, as shown in \citet{Harnois2020}.

Next, in order to aid the physical interpretation of how the WL peak abundances (Fig.~\ref{fig:PA}) depend on the four parameters,  $\Omega_{\rm m}$, $S_8$, $h$ and $w_0$, we  use the \cosmoslics{} data to train a GP emulator as discussed in \ref{sec:Emulation} and present the emulated peak abundances in the \cosmoslics{} parameter space. In Fig.~\ref{fig:PA vary 1 param} we present these emulator predictions while varying one parameter at a time.

The emulated peak abundances plotted in Fig.~\ref{fig:PA vary 1 param} sample the signal around the test cosmology with parameters [$\Omega_{\rm m}$,$S_8$,$h$,$w_0$] = [$0.3$, $0.$8, $0.7$, $-1.0$]. Each sub-panel contains curves where one parameter is varied above and below these values. The bottom row of sub panels shows the ratio of the curves relative to the test cosmology. $1\sigma$ standard errors measured from the $50$ \cosmoslics{} realisations are included on the test cosmology, shown by the shaded blue region. Finally, the $\nu$ range plotted here is slightly narrower than that presented in Fig.~\ref{fig:PA}, since we are now showing the $\nu$ range that will be used to forecast the peak abundance constraints, which is $\nu \in [0,6]$. We do not use peaks with $\nu>6$ for two reasons. First, due to a low number density, the sample variance in this regime is very high. Second, as previously mentioned, this regime is significantly affected by uncertainties such as intrinsic alignments and baryonic physics. We have also performed tests using a higher upper limit of $\nu<8$, and find that it has nearly no impact on our results. Therefore, using $\nu<6$ allows us to stay within the regime that is less affected by the aforementioned uncertainties, whilst still encapsulating maximal cosmological information.

Panel A shows how the peak abundance depends on $\Omega_{\rm m}$. Note that since $S_8 = \sigma_8\big({\Omega_{\rm m}}/{0.3}$\big)$^{0.5}$, $\sigma_8$ increases when $\Omega_{\rm m}$ is reduced (and vice versa), in order for $S_8$ to remain constant. Increasing $\Omega_{\rm m}$ (with $S_8$ and the other parameters held constant) reduces the abundance of WL peaks with amplitudes $\nu < 2$ but increases the abundance of peaks with amplitudes $\nu > 2$, relative to the fiducial case. This is because a higher $\Omega_{\rm m}$ increases the matter content of the universe, which allows dark matter haloes to grow more massive, increasing their lensing signal and the resulting peak amplitudes. The opposite behaviour can be seen when $\Omega_{\rm m}$ is reduced relative to the fiducial case, with more peaks below $\nu = 2$ and fewer above. There is an upturn in the peak abundance for $\Omega_{\rm m} = 0.1$ at $\nu \approx 5$, which is due to the fact that $S_8$ is held constant, rather than $\sigma_8$.

Panel B shows the peak abundance for different $S_8$ values. The results presented in this sub-panel vary more strongly compared to all other sub-panels, verifying that the peak abundance is the most sensitive to $S_8$ of all the parameters studied here. Similar to the behaviour seen for $\Omega_{\rm m}$, increasing $S_8$ reduces the number of small peaks below $\nu \approx 2.7$, but increases the number of large peaks above this point. The opposite behaviour is seen for decreasing $S_8$. Increasing $S_8$ leads to greater clustering of matter, which will place more haloes closer together. The increased mass along an overdense line of sight translates into a greater lensing signal, which produces more peaks of higher amplitudes. This also reduces the number of small peaks since fewer haloes are in isolation which would produce small peaks. 

Panel C shows how the peak abundance changes with $h$. Peaks with amplitudes $\nu < 3$ are mostly unaffected, however there is a small amount of sensitivity to $h$ at the high $\nu$ end, where increasing $h$ slightly increases the number of high peaks and vice versa. This result is not entirely surprising: from Eq.~\eqref{eq:conv source} we can see that the dependencies of $h$, or equivalently $H_0$, cancel out, because the comoving distance can be written as
\begin{equation}
    \chi(z) = \frac{c}{H_0}\int\frac{1}{E(z')}dz'\, ,
\end{equation}
where $E_{\Lambda{\rm CDM}}(z)$ is defined as
\begin{equation}\label{eq:E_LCDM}
E_{\Lambda{\rm CDM}}(z)\equiv \frac{H_{\Lambda{\rm CDM}}(z)}{H_0} = \Omega_{\rm m}(1+z)^3+1-\Omega_{\rm m}\, ,
\end{equation} 
for a flat $\Lambda{\rm CDM}$ cosmology (as is the case of our fiducial cosmology), and is independent of $h$. This means that the $H_0$ factors in the pre-factor, and the $\chi$ and $d\chi$ terms of Eq.~\eqref{eq:conv source} cancel out, so that the only dependence on $H_0$ in $\kappa$ would come through the matter density contrast $\delta$. In the linear-perturbation regime, the evolution of $\delta$ can be expressed in the linear growth factor $D_+$, which for a flat $\Lambda{\rm CDM}$ cosmology is given by the following solution:
\begin{equation}
    D_+(z) = E_{\Lambda{\rm CDM}}(z)\left[\int^\infty_0\frac{(1+z')}{E_{\Lambda{\rm CDM}}^3(z')}dz'\right]^{-1}\int^\infty_z\frac{(1+z')}{E_{\Lambda{\rm CDM}}^3(z')}dz',
\end{equation}
where the term in the bracket offers the normalisation to ensure that $D_+(z=0)=1$ as per the usual convention---this suggests that for flat $\Lambda{\rm CDM}$ models with the same $\Omega_{\rm m}$, $\sigma_8$ and $D_+$, $\kappa$ is independent of $h$. However, we remark that the above argument only applies to a universe with strictly no radiation. In practice, increasing $h$ with $\Omega_{\rm m}$ fixed would mean that the physical matter density today, $\Omega_{\rm m}h^2$, increases, which brings the matter-radiation equality to higher redshift. Since the growth of matter perturbations is slower during radiation domination but faster during matter domination, this means that \textit{small-scale} matter perturbations experience a stronger growth in the case of a larger $h$, and thus it requires a \textit{lower} value of $A_s$ (the amplitude of the primordial power spectrum) in order to reach the desired $\sigma_8$. Consequently, the matter clustering on \textit{large scales}---e.g., at $k$ smaller than $\simeq0.01h{\rm Mpc}^{-1}$, which corresponds to the horizon scale at matter-radiation equality---will indeed be \textit{weaker}. Actually, a more detailed calculation shows that, when comparing the cases of $h=0.9$ and $0.7$ (with $\Omega_{\rm m}$ and $\sigma_8$ fixed), the late-time matter power spectrum $P(k)$ is higher (lower) in the latter than in the former for $k\gtrsim0.1h{\rm Mpc}^{-1}$ ($k\lesssim0.1h{\rm Mpc}^{-1}$). This will have nontrivial implications for the peak 2PCFs as we shall see shortly. Nevertheless, for the peak abundance, the most relevant scales are $k\simeq0.1$--$1h{\rm Mpc}^{-1}$, where the cases of $h=0.7$ and $0.9$ have similar matter clustering amplitudes, which is slightly higher for larger $h$: as this $k$ range corresponds to the sizes of halo-forming regions, this is consistent with the high-$\nu$ behaviour of the middle right panel.

Panel D shows the peak abundances with varying $w_0$. Similar to $h$, the peak abundance does not appear to be very sensitive to changes in $w_0$, but increasing $w_0$ does indeed create slightly more low-$\nu$ peaks and fewer high-$\nu$ peaks compared to the fiducial case, and vice versa. A different dark energy equation of state can change the expansion rate, therefore affecting the comoving distances, the lensing kernel in Eq.~\eqref{eq:conv source}, and the growth rate of matter perturbation $\delta$. The physics underlying the qualitative behaviours shown in these panels is actually complicated and quite interesting. Usually, a more negative $w_0$, e.g., $w_0<-1.0$, implies an increase of the dark energy density with time and therefore (for the same matter density) a faster transition from the phase of decelerated expansion to an accelerated one dominated by dark energy, compared to standard $\Lambda$CDM. But given that we fix $h$ and therefore $H_0$, at $z>0$ the expansion rate is actually slower than the fiducial $\Lambda$CDM model, because the density of dark energy in this case decreases with redshift, and so at $z>0$ the total density of matter and dark energy is smaller than in $\Lambda$CDM. More explicitly, we have $E_{w_0}(z) \leq E_{\Lambda{\rm CDM}}(z)$ for $w_0<-1$, where
\begin{equation}\label{eq:E_DE}
    E_{w_0}(z) = \frac{H_{w_0}(z)}{H_0} = \Omega_{\rm m}(1+z)^3 + \left[1-\Omega_{\rm m}\right](1+z)^{3(1+w_0)},
\end{equation}
which reduces to Eq.~\eqref{eq:E_LCDM} when $w_0=-1$. Because the dark energy in our simulations is assumed to be non-clustering, the only effect of varying $w_0$ is to modify the background expansion history, which leads to a scale-\textit{independent} change in the linear matter clustering, $P(k)$. It may seem that, since $w_0=-1.5$ leads to a slower expansion, it will increase matter clustering. While this is true, we have to note that in this comparison we have fixed $S_8$ (and equivalently $\sigma_8$) today in all three cases, and so this means that, in order to have the same $\sigma_8$ at the present day, the primordial power spectrum must be lower in the $w_0=-1.5$ case (we have explicitly checked this using \textsc{camb}). At an initial thought, this seems to suggest that this model predicts less structure formation than $\Lambda$CDM (until $z=0$), which is against the results of Fig.~\ref{fig:PA vary 1 param}. However, recall that another effect of having a slower expansion is that the Universe becomes older at $z=0$, and distances to the same redshift become larger; the latter, in particular, means that in between the observer and the source(s) there would be more volume, and more structures such as large dark matter haloes. Since these haloes produce the high-$\nu$ peaks, the net effect can be a larger abundance of such peaks. Fig.~\ref{fig:PA vary 1 param} indeed confirms that the two competing effects --- the decrease of large-scale structure due to the lowered primordial power spectrum and the larger volume between the observer and a fixed source redshift --- give rise to a higher peak number count at $\nu\gtrsim2.3$. For low peaks, the total effect is less clear-cut, since a peak is defined as a pixel in the convergence map with a higher $\kappa$ value than all its neighbouring pixels, and both the central and the neighbouring pixels could be affected by chance alignments of small dark matter structures; Fig.~\ref{fig:PA vary 1 param} shows that at $\nu\lesssim2.3$ the peak abundance is smaller in $w_0=-1.5$ than in $\Lambda$CDM. The behaviour of the $w_0=-0.5$ model can be similarly explained. 

%\JHD{[I fear you may loose the reader here, but you could actually clarify that the first effect (linear growth) explains the decrease of peaks observed for small peaks, which are only affected by the chance alignment of the linear LSS, whereas the second effect (non-linear growth) affects the larger peaks, which probe the 1-halo term?]} \YC{\bf (I agree with Joachim about readers may get lost here. I guess the main message to send to the readers is that the abundance respond more sensitively to growth parameters such as $\Omega_m$ and $\sigma_8$, and less sensitively to distance parameters (,which mainly affects the background expansion). This is qualitatively expected, as also shown by the examples in Fig.~3. I am not sure if you need all the detail reasoning for those specific examples.)} \Baojiu{[I think the above explanation is a useful recap of the underlying physics, which is more complicated, so I think it will be beneficial to keep the text (recall that the referee of the void paper asked explicitly for explanations of similar behaviours in that paper). But I have slightly extended it to make the point Joachim mentioned more explicit.]}

Figs.~\ref{fig:PA} and \ref{fig:PA vary 1 param} show that the peak abundance is mostly sensitive to changes in $\Omega_{\rm m}$ and $S_8$, and less so to $h$ and $w_0$, with sensitivity to cosmology coming from both the high and low amplitude peaks.

\subsection{Weak lensing peak two-point correlation function}\label{subsec:peak 2PCF}

\begin{figure*}
    \centering
    \includegraphics[width=1.5\columnwidth]{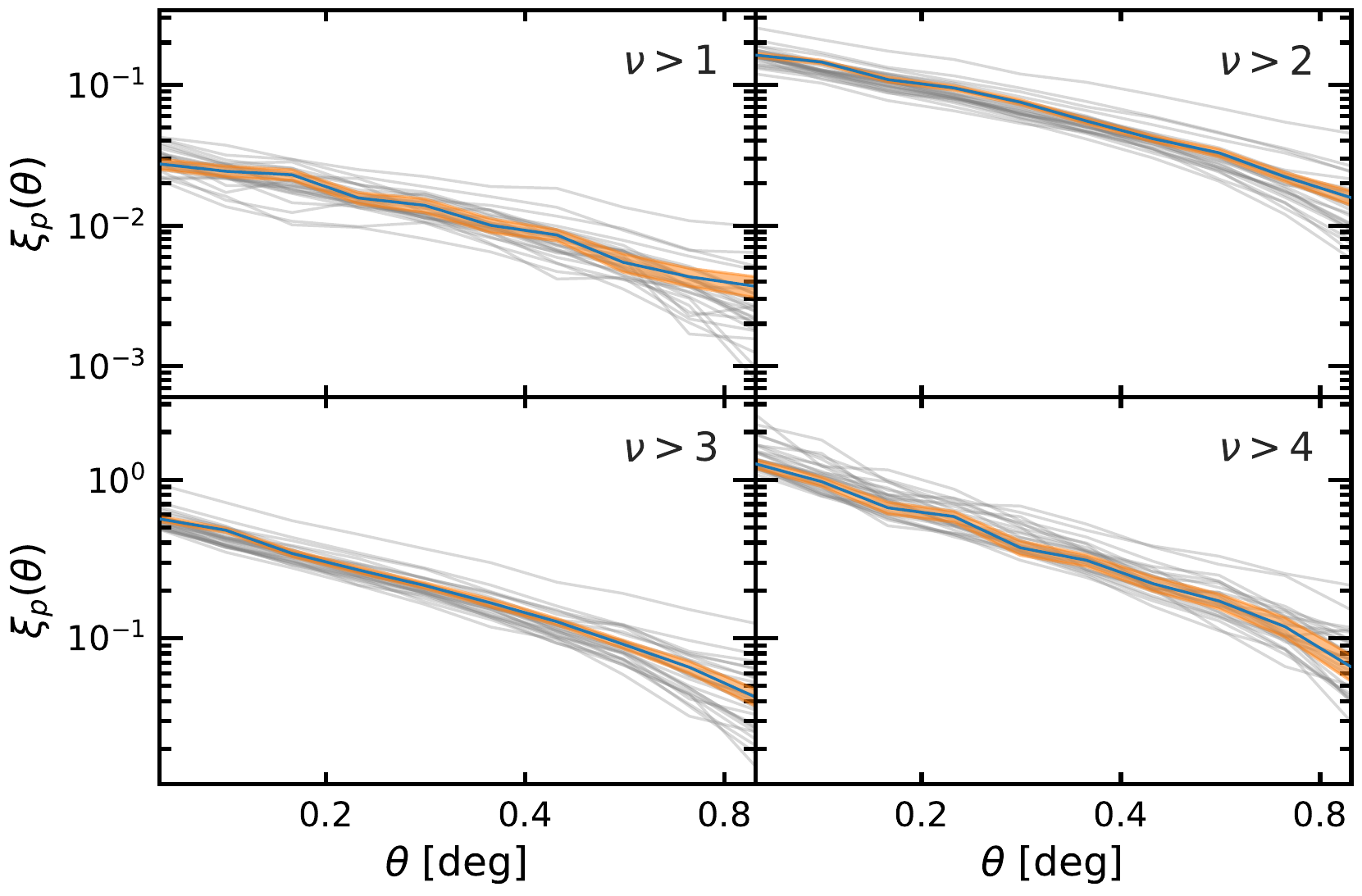}
    \caption{(Colour Online) The WL peak 2PCF, where each subpanel corresponds to the 2PCF of a different peak catalogue. The various peak catalogues (and hence their 2PCF) only contain peaks with amplitudes $\nu > 1$ (top left), $\nu > 2$ (top right), $\nu > 3$ (bottom left) and $\nu > 4$ (bottom right). The curves in each subpanel correspond to the 26 \cosmoslics{} nodes in Fig. \ref{fig:cosmoSLICS nodes}, with the fiducial cosmology plotted in blue. The shaded orange regions show the 1$\sigma$ standard errors measure from the 50 cosmoSLICS realisations. Note the change of $y$-axis range between the upper and lower sub-panels.}
    \label{fig:P2PCFs}
\end{figure*}

\begin{figure*}
    %\centering
    \includegraphics[width=2\columnwidth]{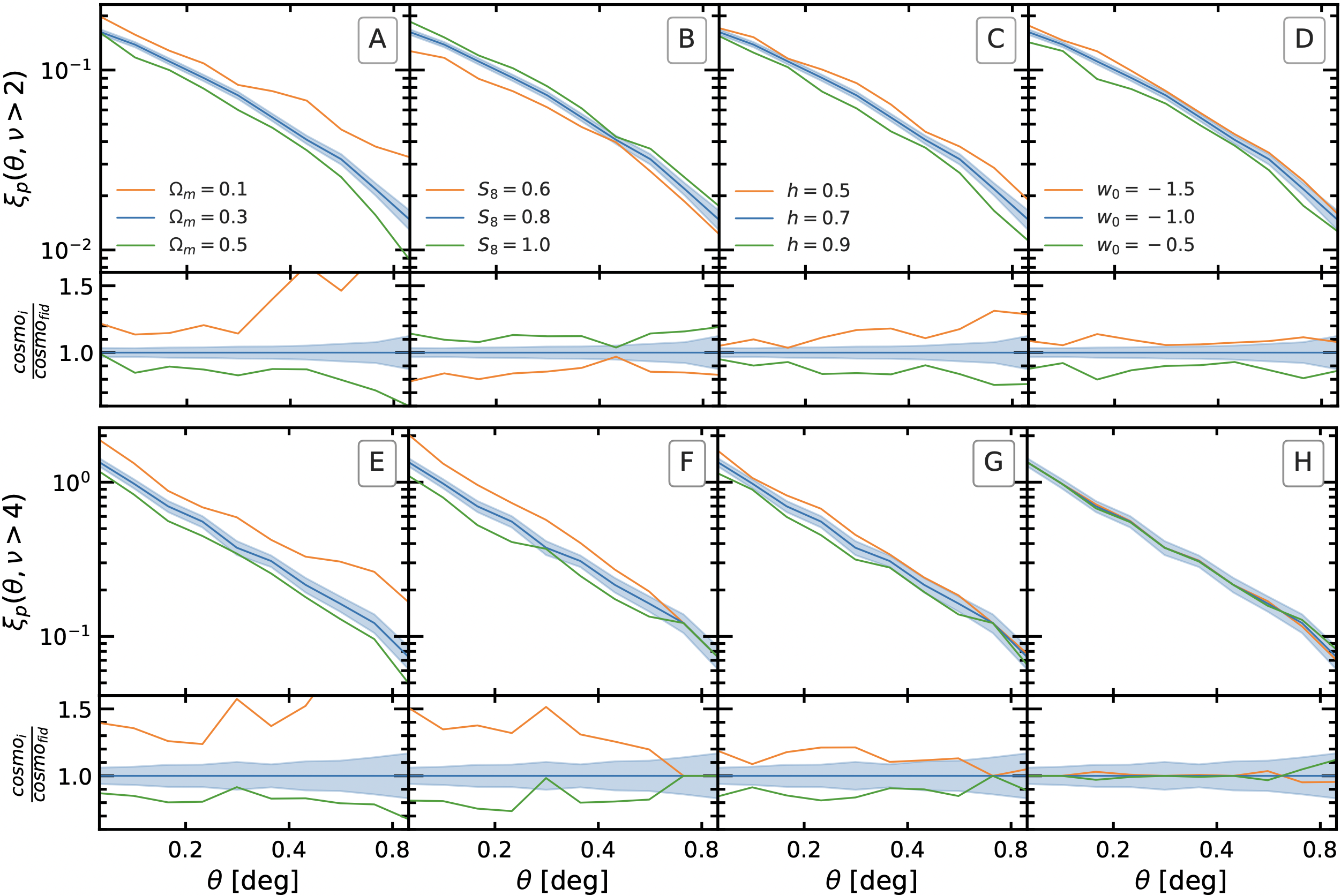}
    \caption{(Colour Online) The emulated peak 2PCF for two different peak catalogues, $\nu>2$ (top section) and $\nu > 4$ (bottom section). The curves correspond to the cosmological parameters [$\Omega_{\rm m}$, $S_8$, $h$, $w_0$] with values [$0.3$, $0.8$, $0.7$, $-1.0$], unless otherwise stated in the sub-panel legends. Each sub-panel corresponds to varying one cosmological parameter at a time, as specified in the legend. The bottom rows in each section show the ratio of the curves relative to the fiducial cosmology. The $1\sigma$ standard errors measured from the $50$ \cosmoslics{} realisations are shown for the fiducial cosmology by the shaded blue region.}
    \label{fig:P2PCFs vary 1 param}
\end{figure*}

In order to measure the WL peak 2PCF, we first remove peaks with amplitudes below a given $\nu$ threshold, and then repeat this step with different $\nu$ thresholds in order to create multiple peak catalogues. This procedure is motivated by the following factors. 

First, as discussed in Section \ref{subsec:peak abubdance}, a significant fraction of the WL peak population is noise-dominated (low $\nu$), suggesting that their spatial distribution (to which the 2PCF is sensitive) may not contain useful cosmological information. We have tested this assertion, and found that the WL peak 2PCF measured using peaks of all heights has a very low amplitude and exhibits only small variations between the 26 \cosmoslics{} nodes. This indicates that when no distinction is made based on peak heights, the average clustering of WL peaks is close to that of a randomly distributed sample. Therefore, in order to extract useful information on the clustering of WL peaks, we must first remove the low-amplitude noise-dominated peaks. We have tested a range of $\nu$ thresholds, and use two criteria to determine the best threshold. First, the amplitude of the 2PCF measurement must be sufficiently high so that the clustering signal is not noise dominated. Second, the variance between 2PCFs for different cosmological parameters must be larger than the variance with which the fiducial 2PCF is measured. This ensures that any cosmological information will not be lost to noisy measurements. We found that both these criteria can be met with a threshold as low as $\nu = 1$.

Second, varying the $\nu$ threshold and using multiple WL peak catalogues produces multiple WL peak 2PCF measurements. The change in the 2PCF as the $\nu$ threshold changes is also sensitive to the underlying cosmology, so we expect that the different 2PCF measurements will contain complementary information to each other, so that combining these measurements will yield tighter cosmological parameter constraints.

Fig.~\ref{fig:P2PCFs} shows the 2PCFs for four WL peak catalogues with $\nu > 1$ (top left), $\nu > 2$ (top right), $\nu > 3$ (bottom left) and $\nu > 4$ (bottom right). The 2PCFs for the 26 \cosmoslics{} nodes are plotted, with the fiducial cosmology plotted in blue, and all other cosmologies plotted in grey. 

The 2PCF measurements for the $\nu > 1$ catalogue have the lowest amplitude. As the $\nu$ threshold increases, so does the amplitude of the 2PCF for all cosmologies, indicating that the high $\nu$ peaks are more clustered than the low $\nu$ peaks. Both the gradient and the amplitude of the 2PCF change as the $\nu$ threshold increases, however the changes in amplitude appear to be the most dominant feature. This can be explained by the relationship between WL peaks and dark matter haloes -- more massive haloes are known to be more strongly biased and clustered, because they form from higher density peaks of the primordial density field.

Similar to Section \ref{subsec:peak abubdance}, we use the \cosmoslics{} data from Fig. \ref{fig:P2PCFs} to train a GP emulator as discussed in \ref{sec:Emulation}, and present emulated peak 2PCFs in the \cosmoslics{} parameter space by varying one parameter at a time. The results are plotted in Fig. \ref{fig:P2PCFs vary 1 param}. The bottom row in each section shows the ratio relative to the test cosmology. The $1\sigma$ standard errors measured from the $50$ \cosmoslics{} realisations are included for the fiducial cosmology and are shown by the shaded blue region. The top and bottom sections of Fig. \ref{fig:P2PCFs vary 1 param} shows results for the 2PCF of peaks with $\nu > 2$ and $\nu > 4$ respectively. We choose to show results for $\nu > 2$ rather than $\nu > 1$ since, as we will see in Fig. \ref{fig:P2PCF contours}, $\nu > 2$ gives stronger parameter constraints than the $\nu>1$ case.

Panel A shows the emulated 2PCF for $\nu > 2$ varying only $\Omega_{\rm m}$. Increasing $\Omega_{\rm m}$ has less of an impact on the 2PCF for small $\theta$ compared to large $\theta$, effectively steepening the curve relative to the fiducial case by a small amount. When decreasing $\Omega_{\rm m}$, the above behaviour is mirrored, except the overall magnitude of the change is larger compared to the case where $\Omega_{\rm m}$ is increased. This shows that $\Omega_{\rm m}$ dictates the gradient of the 2PCF, and appears to be more sensitive to low $\Omega_{\rm m}$ values. It might seem counter-intuitive that a model with smaller $\Omega_{\rm m}$ would predict a stronger clustering for WL peaks, but we note again that here $S_8$ has been fixed when $\Omega_{\rm m}$ is being varied, so that a smaller $\Omega_{\rm m}$ corresponds to a larger $\sigma_8$, and the latter means there is more matter clustering.

Panel E shows the peak 2PCFs for $\nu > 4$ for the same $\Omega_{\rm m}$ values. Similar to the $\nu>2$ case, increasing $\Omega_{\rm m}$ decreases the 2PCF amplitude and vice versa, and the behaviour relative to the fiducial case is asymmetric, where changing $\Omega_{\rm m}$ by a fixed amount in either direction has a larger impact on the amplitude when $\Omega_{\rm m}$ is decreased, suggesting that the $\nu > 4$ catalogue is also more sensitive to small $\Omega_{\rm m}$. However, compared to the $\nu > 2$ case, there appears to be slightly less change to the overall slope of the 2PCF as $\Omega_{\rm m}$ is varied.

Panel B is the same as the previous panels except $S_8$ is varied in this case. The figure shows that changes to $S_8$ affect the amplitude of the 2PCF, where lowering $S_8$ lowers the 2PCF amplitude since it corresponds to a smaller $\sigma_8$ (remember that $\Omega_m$ is held constant here) and therefore less clustering of matter. Increasing $S_8$ by the same amount increases the amplitude relative to the fiducial case, but the magnitude of the change is slightly smaller compared to the decreased $S_8$ case. Panel F shows the $\nu >4$ 2PCF for the same three $S_8$ values. The overall trend here is the opposite to the $\nu>2$ case. Initially it seems counter-intuitive that higher $S_8$ values would lead to a lower clustering amplitude; however, as shown by Fig.~\ref{fig:PA vary 1 param}, the abundance of peaks is also larger for this catalogue. Therefore, when $S_8$ increases, the number of peaks with $\nu > 4$ increases, meaning that smaller maxima in the primordial density field---which are less biased and hence less clustered tracers of the matter density field---end up contributing to this peak catalogue, and so the clustering of the peaks decreases and vice versa. 

%\JHD{[I don't understand that last argument. ]} \Baojiu{[For Joachim: The argument here is actually quite simple:  Fig.~\ref{fig:PA vary 1 param} shows that for a larger $S_8$ you create a lot more higher peaks, which means here you are looking at the clustering of more (initial) density field peaks. This is similar to looking at dark matter haloes with a higher number density and lower mass cutoff, and you will get a lower clustering than another halo catalogue with a higher mass cutoff and lower number density. This happens quite often in modified gravity, where the halo clustering (for haloes of the same mass cutoff) is lower than in $\Lambda$CDM despite the enhanced gravity, because smaller initial density peaks can be `promoted' to go above the chosen mass threshold by the enhanced gravity, hence reducing the overall clustering.]}

Panel C shows how the 2PCF for the $\nu > 2$ catalogue depends on $h$. The 2PCF appears to be sensitive to changes in $h$, where increasing $h$ decreases its amplitude, and vice versa. This observation is actually consistent with the discussion above about the physical impact of varying $h$---with $\Omega_{\rm m}$ and $\sigma_8$ fixed---on matter clustering: increasing $h$ from $0.7$ to $0.9$ weakens the late-time matter clustering at $k\lesssim0.1h{\rm Mpc}^{-1}$, and these are the scales most relevant for the peak clustering (which is expected to trace the dark matter clustering) as well. Unlike the behaviour seen for $\Omega_{\rm m}$ and $S_8$, changing $h$ by a fixed amount in either direction appears to change the 2PCF amplitude by an equal amount. Panel G shows the $\nu > 4$ 2PCF for the same three $h$ values. As for the case of $\nu>2$, the 2PCF amplitude increases when $h$ decreases and vice versa. Indeed, the impacts of varying $h$ are similar for both the $\nu>2$ and the $\nu>4$ catalogues, except the $\nu > 2$ case appears to be more sensitive to $h$ at large $\theta$.

Panel D shows the $\nu > 2$ 2PCFs for different values of $w_0$. Increasing $w_0$ decreases the amplitude and vice versa, with no apparent changes to the gradient. This behaviour also appears to be symmetric relative to the fiducial cosmology, similar to that seen for $h$, and unlike $\Omega_{\rm m}$ and $S_8$. Panel H is the same but shows the $\nu > 4$ 2PCF, which appears to have little sensitivity to changes in $w_0$. In both catalogues, we think that the physical reason underlying the $w_0$ dependence is the same as in the case of peak abundance. Although naively it seems that the case of $w_0=-1.5$---which has faster growth of structures at late times and hence requires a lower initial power spectrum amplitude to achieve the same $\sigma_8$ or $S_8$ at $z=0$---should predict less matter clustering during the entire lensing kernel and so lead to a lower amplitude of the peak 2PCF, we note that this model also covers a larger volume for the same redshift range and therefore receives contribution from a greater number of massive haloes. Also, the peak 2PCF is a projection effect, and the projection depth is larger for the case of $w_0=-1.5$, which leads to a larger line-of-sight integration. These different effects compete with each other and can have cancellations, which may explain why for the $\nu>4$ catalogue there is almost no dependence on $w_0$ (also note that the same $\nu>4$ peak height threshold can lead to different peak populations for the different models, which could also have an impact on the peak correlation).

Comparing the bottom row ($\nu>4$) to the top row ($\nu > 2$), we see that the amplitudes of the 2PCFs are all higher. Given that fewer tracers are used for the $\nu > 4$ measurements, the errors on these curves should be larger. This will be in direct competition with any increased sensitivity to the cosmological parameters relative to the $\nu > 2$ case. Nonetheless, separate catalogues still contain complementary information to each other regardless of which factors wins out, as we will show later.

\section{parameter constraints forecast}
\label{sec:param constraints forecast}

In this section we present parameter constraint forecasts for the statistics studied in Section \ref{sec:weak lensing peak statistics}. 

\begin{figure*}
    \centering
    \includegraphics[width=2\columnwidth]{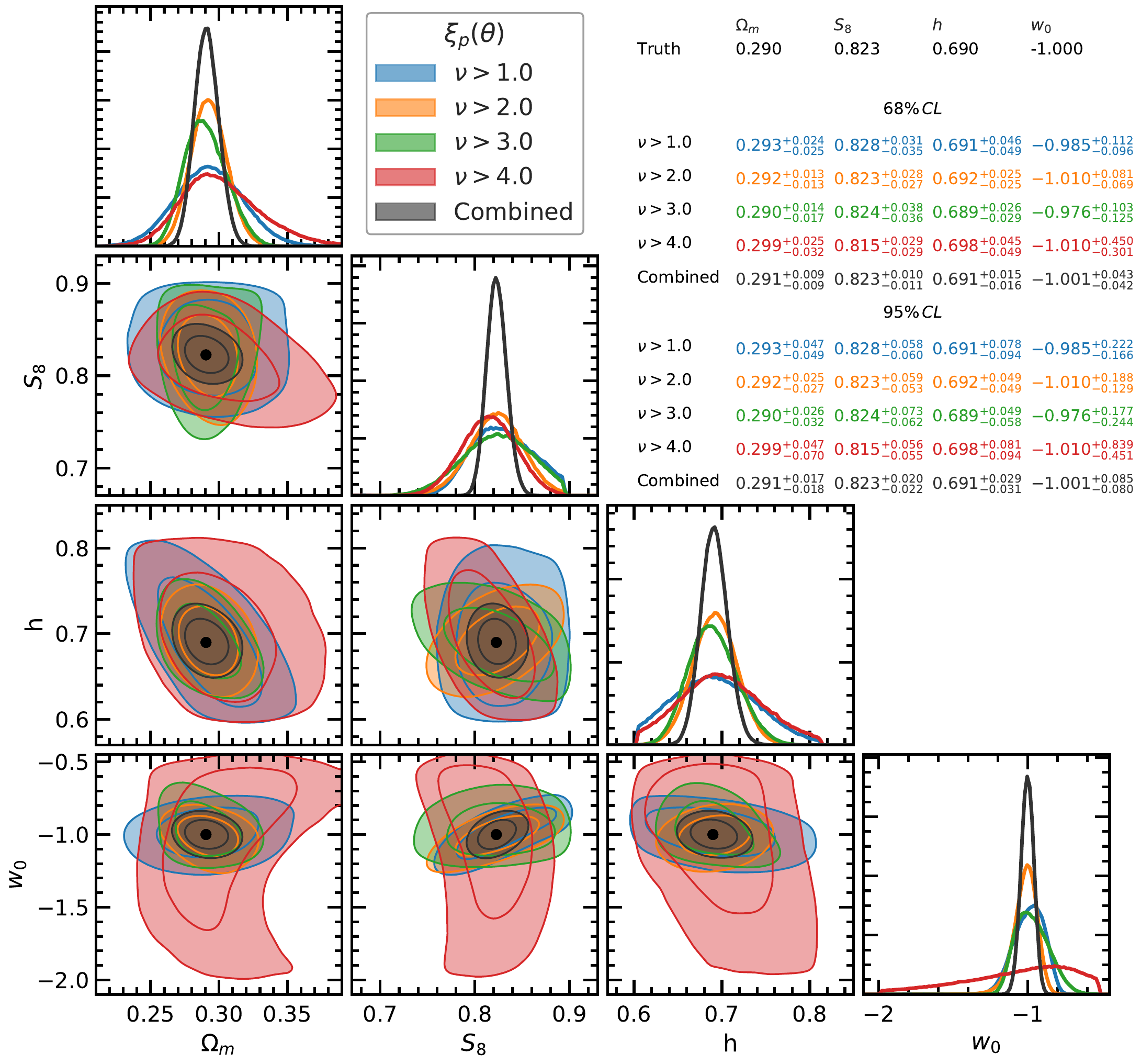}
    \caption{(Colour Online) Constraint forecasts on cosmological parameters measured from the WL peak 2PCF. Contours are shown for 2PCFs measured from WL peak catalogues with $\nu > 1$ (blue), $\nu > 2$ (orange), $\nu > 3$ (green) and $\nu > 4$ (red) and the combination of all four catalogues (black). The true cosmological parameter values used to generate the data are indicated by the black point. The diagonal panels show the 1D marginalised probability distribution, and the remaining panels show the marginalised 2D probability contours enclosing the $68\%$ and $95\%$ confidence intervals. The table in the top right shows true parameter values (top) and the inferred parameter values for the different peak catalogues with $68\%$ (upper section) and $95\%$ (lower section) confidence limits. }
    \label{fig:P2PCF contours}
\end{figure*}

Figure \ref{fig:P2PCF contours} shows the parameter constraint forecasts for an LSST-like survey for the WL peak 2PCF. We present constraints for the four WL peak 2PCFs measured from WL peak catalogues with four heights $\nu > 1,2,3$ and $4$, and the combination of all four 2PCFs. The true cosmological parameter values used to generate the data are indicated by the black point. The diagonal panels show the 1D marginalised probability distributions, while the remaining panels show the marginalised 2D probability contours enclosing the 68\% and 95\% confidence intervals. All confidence intervals, along with the true parameter values, are explicitly stated in the table in the top right of the figure.

In general, as the $\nu$ threshold increases, the contour sizes start off large ($\nu >1$), begin to shrink ($\nu > 2$ and $\nu > 3$), and become large again ($\nu > 4$). The shape and orientation of the contours also change significantly as the $\nu$ threshold increases. For example, in the $\Omega_{\rm m}$--$S_8$ plane, the $\nu > 3$ contour is smaller than the $\nu > 4$ contour; however, the two are orthogonal to each other. This behaviour shows that the constraining power of the WL peak 2PCF can be significantly improved when the 2PCFs of multiple peak catalogues are combined. Even in the case of very large contours which fully enclose the contours from lower $\nu$ thresholds, the presence of complementary information between the different 2PCFs is not ruled out. This is because it depends not only on the size, shape and orientation of the contours, but also on the correlation between the contours. This is discussed in detail in Appendix \ref{app:cov}. The benefit to combining multiple peak catalogues is shown by the grey contours, which are significantly smaller than any individual contour in all cases. 

We find that the $\nu > 2$ and $\nu > 3$ peak 2PCFs give the tightest constraints on $\Omega_{\rm m}$, the $\nu > 2$ and $4$ 2PCFs both give the similar constraint on $S_8$, $\nu > 2$ and $3$ are tightest on $h$ and $\nu > 2$ gives the best constraint on $w_0$. It is interesting to note that the constraints on $w_0$ are roughly nine times smaller for the combination of all catalogues compared to $\nu > 4$ alone, indicating that a significant amount of cosmological information is contained in the clustering of low amplitude peaks. 

\begin{figure*}
    \centering
    \includegraphics[width=2\columnwidth]{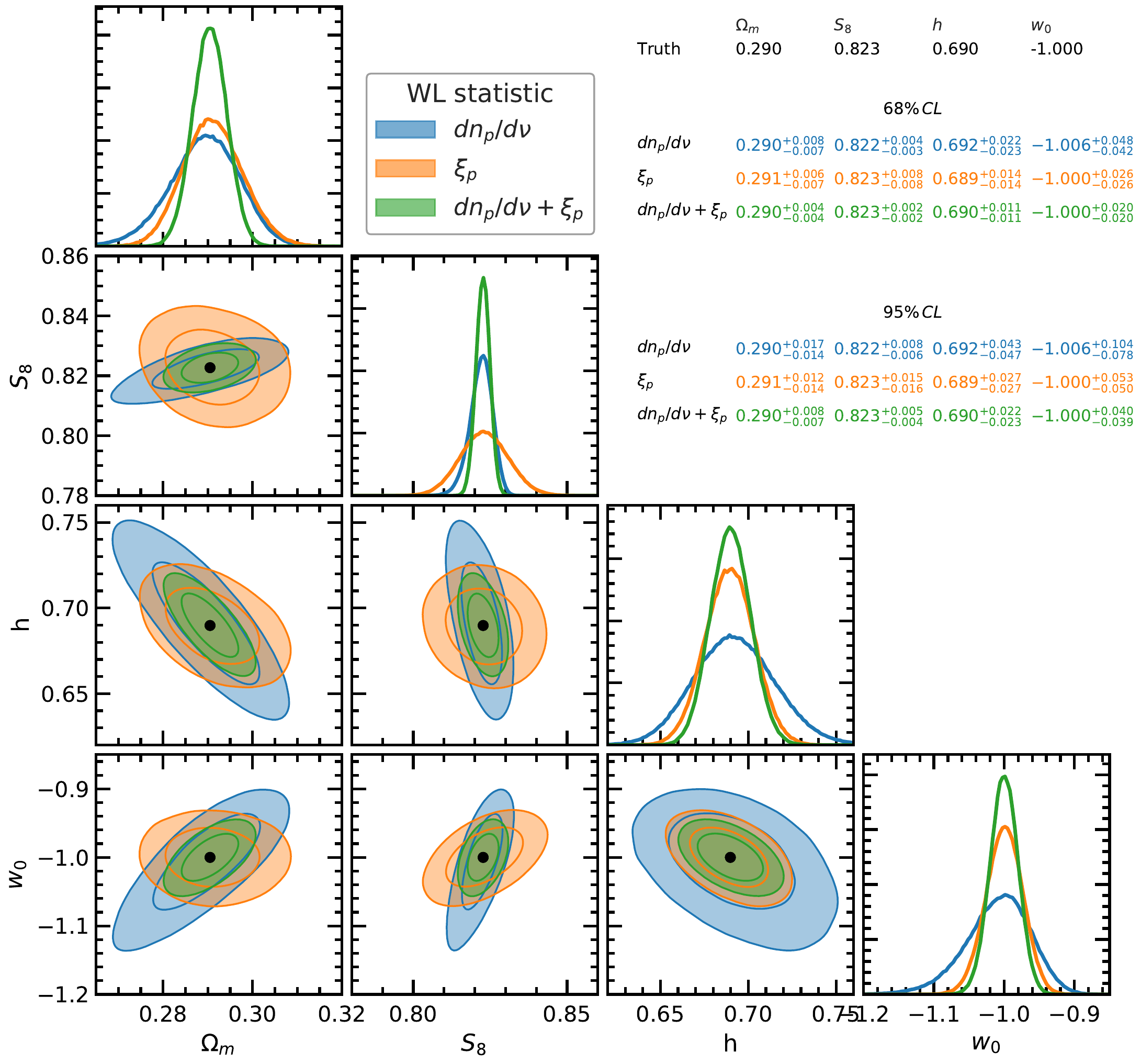}
    \caption{(Colour Online) The same a Fig.~\ref{fig:P2PCF contours} but for the combination of eight peak 2PCFs from peak catalogues with $\nu > 1.0,1.5,...,4.5$ (orange), the peak abundance (blue), and the combination of the two (green).}
    \label{fig:peak contours comparison}
\end{figure*}

Fig.~\ref{fig:peak contours comparison} shows parameter constraint forecasts for the combination of the peak 2PCFs similar to the black contours of Fig. \ref{fig:P2PCF contours}, but now using a finer selection from eight peak catalogues with $\nu > 1.0, 1.5$, $..., 4.5$ in orange. The constraints from the peak abundance for peaks with heights $0 < \nu < 6$ are shown in blue, and the combination of the abundance and 2PCFs are shown in green. We note that when multiple probes are combined, it is important to account for any duplicate information between the probes through the covariance matrix of the data vector, including the cross correlation between the multiple probes. The covariance matrix of all probes studied in this work is presented in Appendix \ref{app:cov} and discussed in detail therein. The orange contour shows how the parameter constraints are improved when the 2PCFs of many more WL peak catalogues are combined: constraints from the combined 2PCFs are much smaller than the best constraints from any individual catalogue (cf.~Fig.~\ref{fig:P2PCF contours}). Increasing the number of catalogues used in the combined case from four to eight, improves the constraints on $\Omega_{\rm m}$ and $S_8$ by roughly $30\%$ and $20\%$ respectively, there is only a small improvement for $h$, and the $w_0$ constraints improve by nearly a factor of two. Potentially one could use a very large number of $\nu$ thresholds, by reducing the $\nu$ increments further, such as $\nu > 1.0, 1.25$, $..., 4.5$. However at some point the 2PCFs from adjacent catalogues become so correlated that there is no gain in extra information, and this significantly increases the length of our data-vector. Our choice is a compromise between having several $\nu$ bins that span a range of peak heights while keeping a small enough data vector to calculate accurate covariance matrices.

The complementarity between 2PCFs with different thresholds, including the potential for degeneracy breaking, is the main factor that contributes to the strong peak 2PCF constraints, when many catalogues are used. In the following paragraphs, we provide a physical interpretation for this behavior.

There have been several studies on the structures that produce weak lensing peaks \citep[][]{Yang2011,J.Liu2016,Wei2018}. These studies show that in general, medium-height peaks correspond to chance alignments of multiple haloes along the line of sight, and high amplitude peaks are often associated with single clusters along the line of sight. Peaks created by LSS projections will contain different cosmological information to those created by single clusters. For the peak abundance, this can be seen in the covariance matrix (Fig. \ref{fig:covariance}, also studied in \citealt{Martinet2018}) where there is anti-correlation between low and high peaks. There is a similar behavior in the peak 2PCF covariance matrix in Fig. \ref{fig:covariance peaks}, where there is very little correlation between the $\nu > 1$ and $\nu > 4$ catalogues. However this feature is less pronounced, due to adjacent catalogues sharing a large fraction of the same peaks.

Because peaks of different amplitudes are produced by different types of dark matter structures, the 2PCF of each WL peak catalogue probes slightly different regimes of the dark matter distribution. By combining multiple peak catalogues with different peak height thresholds, we essentially add additional information about the peak height to the 2PCF.

As shown by the table in Fig.~\ref{fig:peak contours comparison}, the peak abundance and peak 2PCF provide similar constraints on $\Omega_{\rm m}$, however the constraints on $S_8$ are twice as strong for the peak abundance compared to the peak 2PCF: in the $\Omega_{\rm m}$--$S_8$ plane, the peak abundance contour is significantly tighter than the peak 2PCF contour in the $S_8$ direction. When the two probes are combined, there is an overall improvement on the $\Omega_{\rm m}$ and $S_8$ constraints by a factor of two, relative to the peak abundance alone. This leads to a good overall improvement in the $\Omega_{\rm m}$--$S_8$ plane when the peak abundance and 2PCF are combined, as shown by the green contour. 

The peak 2PCF is able to constrain both $h$ and $w_0$ with greater accuracy than the peak abundance. There also appears to be some orthogonality between the abundance and 2PCF constraints in the $h$--$S_8$ and $w_0$--$\Omega_{\rm m}$ planes. In the $w_0$--$h$ plane, the parameter constraints are dominated by the peak 2PCF contours, while the peak abundance contours are significantly larger than the former. This indicates that the peak 2PCF offers a great deal of complementary information to the peak abundance, and combining the two probes can significantly improve constraints in the four dimensional parameter space studied here. The behaviour of the constraints from peak abundance and 2PCF, especially those on $h$ and $w_0$, are consistent with the observations we made above for Figs.~\ref{fig:PA vary 1 param} and \ref{fig:P2PCFs vary 1 param}.

\begin{figure*}
    \centering
    \includegraphics[width=2\columnwidth]{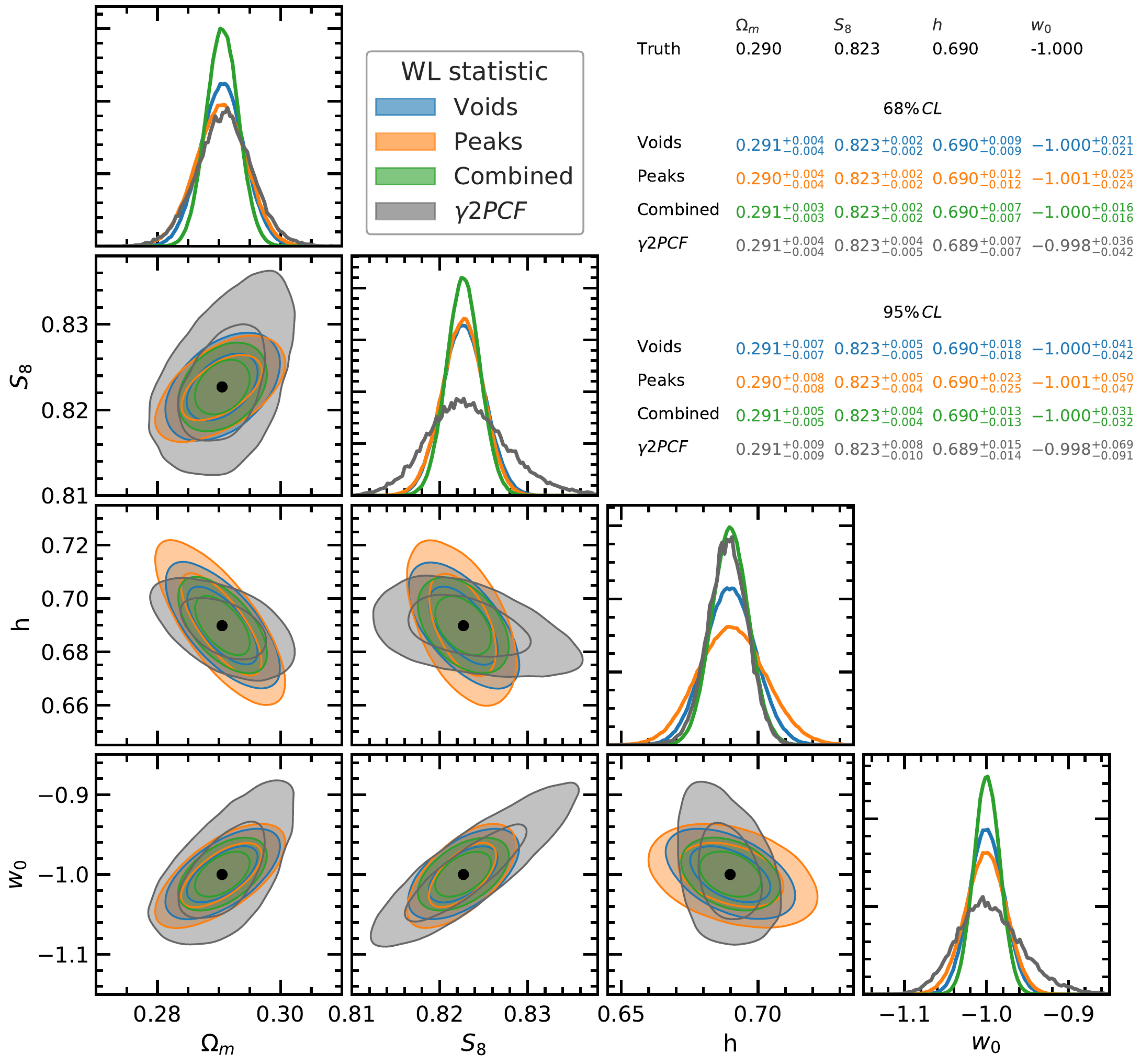}
    \caption{(Colour Online) The same as Fig. \ref{fig:P2PCF contours}, but for void statistics presented in \citep{Davies2020b} (blue), peak statistics (orange) and the combination of peak and void statistics (green). Both the peak and void statistics use the combination of the $\nu > 1,2,3$ and $4$ catalogues. Shear 2PCF forecasts are shown in grey.}
    \label{fig:peak and void contours comparison}
\end{figure*}

In Fig.~\ref{fig:peak and void contours comparison} we also introduce the parameter constraint forecasts for WL voids from \cite{Davies2020b}, which are measured with the same methodology and specifications used in this work, to compare the constraining power of these two different probes. This is important since the voids studied in \cite{Davies2020b}, which were found to be a promising void definition \citep{Davies2020}, are identified as underdense regions in the distribution of WL peaks. This means that the properties of WL voids are likely correlated with the number and clustering of peaks, and we need a joint analysis to reveal the amount of complementary information contained in the two probes. The forecasts from the WL voids (blue), making use of both their abundances and tangential shear profiles, are compared to the WL peak forecasts (orange), which combine the peak abundance and peak 2PCF. We note that both the void and peak contours presented here are for the combination of the $\nu > 1,2,3$ and $4$ peak catalogues (excluding the peak abundance which does not combine multiple catalogues). This is to provide a fair comparison between the voids and the peaks. In principle, the void contours could be measured for the eight catalogues used for the orange peak 2PCF contours in Fig. \ref{fig:peak contours comparison}, however this would cause our data vector to become too large, even for our high number of SLICS realisations. 

Overall, both the peaks and voids are able to measure the four cosmological parameters with similar accuracy. The voids provide notably tighter measurements of $h$ and $w_0$. The void contours are smaller than the peak contours, and follow similar degeneracy directions for all combinations of parameters. The void and peak contours are most similar in the $S_8$--$w_0$ plane, and most distinct in the $w_0$--$h$ plane. When the peaks and voids are combined (green contours), there is a small improvement on the $\Omega_{\rm m}$, $S_8$ and $h$ measurements, and, there is also a reasonable improvement on $w_0$, indicating that the WL peak and void statistics are complementary to each other.

As a comparison, we also include the forecast contours using the standard cosmic shear 2PCFs ($\xi_+$ and $\xi_-$ combined) in grey \citep[for details on how these are measured, see][]{Davies2020b}. For fair comparisons, the cosmological model dependence and covariance matrix for these were both obtained using the same simulation data as used for the peak and void analyses throughout this paper. For $\Omega_{\rm m}$ and $h$, WL peaks or WL voids (or both of them) give similar constraints as the shear 2PCFs; however, for $S_8$ and $w_0$, the former probes actually can place tighter constraints (for this survey specification) by roughly a factor of two, indicating again the benefit of exploring beyond-two-point WL statistics to help maximise the information that can be extracted. In some parameter planes, such as $S_8$--$h$ and $h$--$w_0$, there is a clear orthogonality between the degeneracy directions of the peak/void statistics and the shear 2PCFs.

%\mcc{Marius: In my opinion, judging from the diagonal panels in Fig. 8, $h$ shows the largest improvement, followed by $\Omega_{\rm m}$, while the other 2 parameters show a similar gain when combining peaks and voids. In general, the amount of improvement can be subjective, so we need to be careful how we present it. Maybe we could mention the decrease in errors when combining voids and peaks , e.g. "a small improvement on the $S_8$ and $w_0$ measurement ($\sim XX\%$ decrease in errors) versus a larger improvement in $\Omega_{\rm m}$ and especially in $h$ corresponding to a decrease of respectively $YY$ and $ZZ\%$ in uncertainties."}

\section{Discussion and conclusions}\label{sec:discussion and conclusions}

We have tested the sensitivity of the WL peak statistics to the cosmological parameters $\Omega_{\rm m}$, $S_8$, $h$ and $w_0$ and compared the peak 2PCF to the peak abundance. In order to achieve this, we have trained a Gaussian Process emulator with 26 cosmologies sampled in the 4D parameter space using a Latin hypercube, which we used to predict the peak statistics for arbitrary cosmologies (within the range spanned by the training cosmologies). We have run Markov Chain Monte Carlo samplings from our mock weak lensing data to forecast the accuracy's at which these four parameters can be constrained by a future, LSST-like, lensing survey, using the above WL peak statistics. 

Using the emulators, we have studied the behaviour of the WL peak 2PCF in detail, and %we have 
made connections to the well-established peak abundance. A main feature of our peak 2PCF analysis is that we generate a WL peak catalogue from the entire peak population by introducing a peak height ($\nu$) threshold, below which all peaks are removed, and then vary this threshold to generate multiple catalogues. We then study the behaviour of the WL peak 2PCF of these catalogues as this $\nu$ threshold changes.

In \cite{Marian2013}, it has been shown that the WL peak 2PCF of high-amplitude peaks provides little complementary information to the peak abundance. In this work, we have presented some additional steps that are able to further push the utility of the WL peak 2PCF. These additional steps significantly improve the overall constraining power of WL peaks when the abundance and 2PCFs are combined. First, we study the 2PCF of low-amplitude ($\nu$) peaks, and find that it contains significant cosmological information compared to the 2PCF of high-amplitude peaks. For example, in Fig.~\ref{fig:P2PCF contours} the constraints on $w_0$ are roughly four times stronger for the $\nu > 2$ catalogue compared to the $\nu > 4$ catalogue. Second, we find that the 2PCFs of multiple catalogues are complementary to each other, and when combined, the peak 2PCF can constrain $\Omega_{\rm m}$ with tighter accuracy than the peak abundance, and that it is able to constrain both $h$ and $w_0$ with significantly greater accuracy than the peak abundance alone. We also find that the peak abundance provides constraints that are twice as tight on $S_8$ than the combined peak 2PCF, indicating that in order to fully exploit the cosmological information contained in WL peaks, both their abundance and their clustering should be measured and combined. This is illustrated by the green contours in Fig.~\ref{fig:peak contours comparison}, which show the total constraints from WL peaks in which the abundance is combined with the combined 2PCF from different $\nu$ catalogues. Here, the abundance plus the clustering forecasts are roughly twice as strong as those for either of the individual cases (orange for 2PCF and blue for abundance). When we compare the constraints from the peak abundance plus the peak 2PCF to those from the shear 2PCF, we find that the peaks are able to constrain $\Omega_{\rm m}$, $S_8$ and $w_0$ with greater precision than the shear 2PCF, the most significant improvement is for $S_8$ and $w_0$, which improve by roughly a factor of two. Finally, the information required to measure the peak 2PCF is already present when the peak abundance is measured. Therefore, the addition of the peak 2PCF to any preexisting peak abundance analysis pipeline will require minimal modifications, making the peak 2PCF a very promising probe.

We also include a comparison of the forecasts from WL peaks to the WL voids studied in \cite{Davies2020b}, and find that the combination of the two can improve the constraints on $\Omega_{\rm m}$, $S_8$ and $h$, and can provide significant improvements on the $w_0$ measurements. The WL voids are sensitive to the $N$-point correlation function of peaks \citep{White1979}, and the improved constraints resulting from combining WL peaks and voids shows that three and higher-order correlation functions in the peak distribution contain complementary cosmological information. WL voids are one simple way to access the information contained in the higher order correlation functions of peaks. 

The improved parameter constraints from the combination of all of the peak and void probes presented here can also be explained through their full covariance matrix, which includes cross-correlations between probes, present in Appendix \ref{app:cov}. From the covariance matrix it is clear that many of the probes studied here have a large degree of statistical independence. This leads to complementary information between the probes which yields tighter constraints when the different probes are combined. 

We highlight that the work carried out here applies to the 4D parameter space in Fig.~\ref{fig:cosmoSLICS nodes}, and may change if additional parameters in the $\Lambda{\rm CDM}$ model, such as the spectral index, are included. Our results may also be sensitive to changes in curvature, massive neutrinos or other sources of additional physics. For example in \cite{Davies2019b} we found that the peak abundance is sensitive to the nDGP modified gravity model, and \cite{X.Liu2016} have used the WL peak abundance to constrain $f(R)$ gravity.

We also note that the ray-tracing method used to obtain our WL maps employs some approximations, the most important being the Born approximation. We do not expect this simplified framework to have affected our results. For example, \cite{Hilbert2020} have compared multiple ray-tracing methods, such as our approach versus full techniques run on the fly at the same time as the numerical simulation, and found very good agreement in the WL convergence and shear power spectrum, as well as in the abundance of peaks with heights $\nu<6$. \citeauthor{Hilbert2020} have found some discrepancies between methods in the peak abundance for $\nu>6$, which should not be surprising since such peaks correspond to massive clusters, however such differences are unlikely to affect our results since peaks with such high amplitudes constitute only a very small fraction of the population even for our $\nu>4$ peak catalogue (e.g. see Fig.~\ref{fig:PA}). 
The agreement of different ray-tracing methods when predicting the peak 2PCF remains to be studied. In this work the peak 2PCF is measured on scales larger than 0.1 $\deg$, so any approximations we employed, such as the Born approximation, need to fail on similar scales to have an impact on our measurements.

In addition, the simulations used to construct the emulators for the different WL statistics analysed here are limited in their number of nodes sampled with the Latin hypercube. As the results of this paper suggest, future WL observations can place competitive constraints on the various cosmological parameters, with significantly smaller contours than the current status. As the contours keep shrinking around the best-fit model, improved emulators which can more accurately capture the small effects induced by small variations of parameters will be needed. In the future, it will be necessary to simulate cosmological models sampled using a nested Latin hypercube, or nested Latin hypercubes, to refine the emulators used in this work.

The results presented here may be further improved with the inclusion of tomography. This is the standard approach when using the shear two-point correlation function to measure cosmological parameters, and typically this significantly improves the constraints on $w_0$. Therefore, in a future work, we will test how cosmological parameter constraints can be improved when using WL voids with tomography. 

Finally, in order to use the WL peak 2PCF in observations, it will be important to understand the impact of baryonic physics, the intrinsic alignment of galaxies, and the uncertainty associated with photometric redshifts and shear calibration. It is already established that the WL peak abundance is altered in presence of baryons \citep{Osato2015,Weiss2019,Coulton2019,Fong2019} and intrinsic galaxy alignments \citep{harnois2021}, so it will also be necessary to test how the WL peak 2PCF is affected by these. 
The behaviour of these systematics will be an important factor in determining which peak catalogues can be used to reliably measure the peak 2PCF in observations. When identifying WL peaks in observational data, previous studies \citep[e.g.][]{Kacprzak2016,Martinet2018,Harnois2020} have discarded peaks with $\nu < 0$ and $\nu > 4$, in order to mitigate the impact of these systematics on the peak abundance. We would expect a similar approach to be valid for the peak 2PCF, though that remains to be tested. Additionally, the removal of the high-amplitude peaks that are most affected by these systematics may not change our posterior forecasts by a significant amount, since they also contribute the least to the parameter constraints.

\section*{Acknowledgements}
We thank Ludovic Van Waerbeke for sharing the code used in the weak lensing convergence map reconstruction.

CTD is funded by a UK Science and Technology Facilities Council (STFC) PhD studentship through grant ST/R504725/1. MC is supported by the EU Horizon 2020 research and innovation programme under a Marie Sk{\l}odowska-Curie grant agreement 794474 (DancingGalaxies).
JHD acknowledges support from an STFC Ernest Rutherford Fellowship (project reference ST/S004858/1). BG acknowledges the support of the Royal Society through an Enhancement Award (RGF/EA/181006). YC acknowledges the support of the Royal Society through a University Research Fellowship and an Enhancement Award. BL is supported by an ERC Starting Grant, ERC-StG-PUNCA-716532, and additionally supported by the STFC Consolidated Grants [ST/P000541/1, ST/T000244/1].

The SLICS numerical simulations can be found at http://slics.roe.ac.uk/, while the {\it cosmo}-SLICS can be made available upon request. This work used the DiRAC@Durham facility managed by the Institute for Computational Cosmology on behalf of the STFC DiRAC HPC Facility (www.dirac.ac.uk). The equipment was funded by BEIS capital funding via STFC capital grants ST/K00042X/1, ST/P002293/1, ST/R002371/1 and ST/S002502/1, Durham University and STFC operations grant ST/R000832/1. DiRAC is part of the National e-Infrastructure.

\section{Data Availability}

The data used in this work is available upon request. See \cite{Harnois2019} for more details. 
%%%%%%%%%%%%%%%%%%%%%%%%%%%%%%%%%%%%%%%%%%%%%%%%%%



\bibliographystyle{mnras}
\bibliography{mybib} % if your bibtex file is called example.bib

\begin{thebibliography}{}
\makeatletter
\relax
\def\mn@urlcharsother{\let\do\@makeother \do\$\do\&\do\#\do\^\do\_\do\%\do\~}
\def\mn@doi{\begingroup\mn@urlcharsother \@ifnextchar [ {\mn@doi@}
  {\mn@doi@[]}}
\def\mn@doi@[#1]#2{\def\@tempa{#1}\ifx\@tempa\@empty \href
  {http://dx.doi.org/#2} {doi:#2}\else \href {http://dx.doi.org/#2} {#1}\fi
  \endgroup}
\def\mn@eprint#1#2{\mn@eprint@#1:#2::\@nil}
\def\mn@eprint@arXiv#1{\href {http://arxiv.org/abs/#1} {{\tt arXiv:#1}}}
\def\mn@eprint@dblp#1{\href {http://dblp.uni-trier.de/rec/bibtex/#1.xml}
  {dblp:#1}}
\def\mn@eprint@#1:#2:#3:#4\@nil{\def\@tempa {#1}\def\@tempb {#2}\def\@tempc
  {#3}\ifx \@tempc \@empty \let \@tempc \@tempb \let \@tempb \@tempa \fi \ifx
  \@tempb \@empty \def\@tempb {arXiv}\fi \@ifundefined
  {mn@eprint@\@tempb}{\@tempb:\@tempc}{\expandafter \expandafter \csname
  mn@eprint@\@tempb\endcsname \expandafter{\@tempc}}}

\bibitem[\protect\citeauthoryear{{Aihara} et~al.,}{{Aihara}
  et~al.}{2019}]{Aihara2019}
{Aihara} H.,  et~al., 2019, \mn@doi [\pasj] {10.1093/pasj/psz103}, \href
  {https://ui.adsabs.harvard.edu/abs/2019PASJ...71..114A} {71, 114}

\bibitem[\protect\citeauthoryear{Anderson}{Anderson}{2003}]{Anderson2003}
Anderson T.~W.,  2003, An introduction to multivariate statistical analysis.
Wiley-Interscience

\bibitem[\protect\citeauthoryear{{Asgari} et~al.,}{{Asgari}
  et~al.}{2020}]{Asgari2020}
{Asgari} M.,  et~al., 2020, arXiv e-prints, \href
  {https://ui.adsabs.harvard.edu/abs/2020arXiv200715633A} {p. arXiv:2007.15633}

\bibitem[\protect\citeauthoryear{{Bacon}, {Refregier}  \& {Ellis}}{{Bacon}
  et~al.}{2000}]{Bacon2000}
{Bacon} D.~J.,  {Refregier} A.~R.,   {Ellis} R.~S.,  2000, \mn@doi [Mon. Not.
  Roy. Astron. Soc.] {10.1046/j.1365-8711.2000.03851.x}, 318, 625

\bibitem[\protect\citeauthoryear{{Bartelmann} \& {Schneider}}{{Bartelmann} \&
  {Schneider}}{2001}]{Bartelmann2001}
{Bartelmann} M.,  {Schneider} P.,  2001, \mn@doi [Phys. Rept.]
  {10.1016/S0370-1573(00)00082-X}, \href
  {https://ui.adsabs.harvard.edu/\#abs/2001PhR...340..291B} {340, 291}

\bibitem[\protect\citeauthoryear{{Burger}, {Schneider}, {Demchenko},
  {Harnois-Deraps}, {Heymans}, {Hildebrand t}  \& {Unruh}}{{Burger}
  et~al.}{2020}]{Burger2020}
{Burger} P.,  {Schneider} P.,  {Demchenko} V.,  {Harnois-Deraps} J.,  {Heymans}
  C.,  {Hildebrand t} H.,   {Unruh} S.,  2020, \mn@doi [\aap]
  {10.1051/0004-6361/202038694}, \href
  {https://ui.adsabs.harvard.edu/abs/2020A&A...642A.161B} {642, A161}

\bibitem[\protect\citeauthoryear{{Coulton}, {Liu}, {McCarthy}  \&
  {Osato}}{{Coulton} et~al.}{2019}]{Coulton2019}
{Coulton} W.~R.,  {Liu} J.,  {McCarthy} I.~G.,   {Osato} K.,  2019, preprint,
  \href {https://ui.adsabs.harvard.edu/abs/2019arXiv191004171C} {} (\mn@eprint
  {arXiv} {1910.04171})

\bibitem[\protect\citeauthoryear{{DES Collaboration} et~al.,}{{DES
  Collaboration} et~al.}{2021}]{DES2021}
{DES Collaboration} et~al., 2021, arXiv e-prints, \href
  {https://ui.adsabs.harvard.edu/abs/2021arXiv210513549D} {p. arXiv:2105.13549}

\bibitem[\protect\citeauthoryear{{Davies}, {Cautun}  \& {Li}}{{Davies}
  et~al.}{2018}]{Davies2018}
{Davies} C.~T.,  {Cautun} M.,   {Li} B.,  2018, \mn@doi [Mon. Not. Roy. Astron.
  Soc.] {10.1093/mnrasl/sly135}, \href
  {https://ui.adsabs.harvard.edu/\#abs/2018MNRAS.480L.101D} {480, L101}

\bibitem[\protect\citeauthoryear{{Davies}, {Cautun}  \& {Li}}{{Davies}
  et~al.}{2019a}]{Davies2019}
{Davies} C.~T.,  {Cautun} M.,   {Li} B.,  2019a, \mn@doi [Mon. Not. Roy.
  Astron. Soc.] {10.1093/mnras/stz2157}, \href
  {https://ui.adsabs.harvard.edu/abs/2019MNRAS.488.5833D} {488, 5833}

\bibitem[\protect\citeauthoryear{{Davies}, {Cautun}  \& {Li}}{{Davies}
  et~al.}{2019b}]{Davies2019b}
{Davies} C.~T.,  {Cautun} M.,   {Li} B.,  2019b, \mn@doi [Mon. Not. Roy.
  Astron. Soc.] {10.1093/mnras/stz2933}, \href
  {https://ui.adsabs.harvard.edu/abs/2019MNRAS.490.4907D} {490, 4907}

\bibitem[\protect\citeauthoryear{{Davies}, {Paillas}, {Cautun}  \&
  {Li}}{{Davies} et~al.}{2020a}]{Davies2020}
{Davies} C.~T.,  {Paillas} E.,  {Cautun} M.,   {Li} B.,  2020a, arXiv e-prints,
  \href {https://ui.adsabs.harvard.edu/abs/2020arXiv200411387D} {p.
  arXiv:2004.11387}

\bibitem[\protect\citeauthoryear{{Davies}, {Cautun}, {Giblin}, {Li},
  {Harnois-D{\'e}raps}  \& {Cai}}{{Davies} et~al.}{2020b}]{Davies2020b}
{Davies} C.~T.,  {Cautun} M.,  {Giblin} B.,  {Li} B.,  {Harnois-D{\'e}raps} J.,
    {Cai} Y.-C.,  2020b, arXiv e-prints, \href
  {https://ui.adsabs.harvard.edu/abs/2020arXiv201011954D} {p. arXiv:2010.11954}

\bibitem[\protect\citeauthoryear{Dietrich \& Hartlap}{Dietrich \&
  Hartlap}{2010}]{Dietrich2010}
Dietrich J.~P.,  Hartlap J.,  2010, \mn@doi [Mon. Not. Roy. Astron. Soc.]
  {10.1111/j.1365-2966.2009.15948.x}, 402, 1049

\bibitem[\protect\citeauthoryear{{Fong}, {Choi}, {Catlett}, {Lee}, {Peel},
  {Bowyer}, {King}  \& {McCarthy}}{{Fong} et~al.}{2019}]{Fong2019}
{Fong} M.,  {Choi} M.,  {Catlett} V.,  {Lee} B.,  {Peel} A.,  {Bowyer} R.,
  {King} L.~J.,   {McCarthy} I.~G.,  2019, arXiv e-prints, \href
  {https://ui.adsabs.harvard.edu/abs/2019arXiv190703161F} {}

\bibitem[\protect\citeauthoryear{{Foreman-Mackey}, {Hogg}, {Lang}  \&
  {Goodman}}{{Foreman-Mackey} et~al.}{2013}]{Foreman2013}
{Foreman-Mackey} D.,  {Hogg} D.~W.,  {Lang} D.,   {Goodman} J.,  2013, \mn@doi
  [\pasp] {10.1086/670067}, \href
  {https://ui.adsabs.harvard.edu/abs/2013PASP..125..306F} {125, 306}

\bibitem[\protect\citeauthoryear{{Fu} et~al.,}{{Fu} et~al.}{2008}]{Fu2008}
{Fu} L.,  et~al., 2008, \mn@doi [\aap] {10.1051/0004-6361:20078522}, \href
  {https://ui.adsabs.harvard.edu/\#abs/2008A&A...479....9F} {479, 9}

\bibitem[\protect\citeauthoryear{{Giblin} et~al.,}{{Giblin}
  et~al.}{2018}]{Giblin2018}
{Giblin} B.,  et~al., 2018, \mn@doi [Mon. Not. Roy. Astron. Soc.]
  {10.1093/mnras/sty2271}, \href
  {https://ui.adsabs.harvard.edu/abs/2018MNRAS.480.5529G} {480, 5529}

\bibitem[\protect\citeauthoryear{Habib, Heitmann, Higdon, Nakhleh  \&
  Williams}{Habib et~al.}{2007}]{Habib2007}
Habib S.,  Heitmann K.,  Higdon D.,  Nakhleh C.,   Williams B.,  2007, \mn@doi
  [Phys. Rev. D] {10.1103/PhysRevD.76.083503}, 76, 083503

\bibitem[\protect\citeauthoryear{Hamana, Yoshida  \& Takada}{Hamana
  et~al.}{2004}]{Hamana2004}
Hamana T.,  Yoshida N.,   Takada M.,  2004, \mn@doi [Mon. Not. Roy. Astron.
  Soc.] {10.1111/j.1365-2966.2004.07691.x}, 350, 893

\bibitem[\protect\citeauthoryear{{Hamana}, {Shirasaki}  \& {Lin}}{{Hamana}
  et~al.}{2020}]{Hamana2020}
{Hamana} T.,  {Shirasaki} M.,   {Lin} Y.-T.,  2020, \mn@doi [\pasj]
  {10.1093/pasj/psaa068}, \href
  {https://ui.adsabs.harvard.edu/abs/2020PASJ...72...78H} {72, 78}

\bibitem[\protect\citeauthoryear{{Harnois-D{\'e}raps} \& {van
  Waerbeke}}{{Harnois-D{\'e}raps} \& {van Waerbeke}}{2015}]{Harnois2015}
{Harnois-D{\'e}raps} J.,  {van Waerbeke} L.,  2015, \mn@doi [Mon. Not. Roy.
  Astron. Soc.] {10.1093/mnras/stv794}, \href
  {https://ui.adsabs.harvard.edu/abs/2015MNRAS.450.2857H} {450, 2857}

\bibitem[\protect\citeauthoryear{{Harnois-D{\'e}raps}
  et~al.,}{{Harnois-D{\'e}raps} et~al.}{2018}]{Harnois2018}
{Harnois-D{\'e}raps} J.,  et~al., 2018, \mn@doi [Mon. Not. Roy. Astron. Soc.]
  {10.1093/mnras/sty2319}, \href
  {https://ui.adsabs.harvard.edu/abs/2018MNRAS.481.1337H} {481, 1337}

\bibitem[\protect\citeauthoryear{{Harnois-D{\'e}raps}, {Giblin}  \&
  {Joachimi}}{{Harnois-D{\'e}raps} et~al.}{2019}]{Harnois2019}
{Harnois-D{\'e}raps} J.,  {Giblin} B.,   {Joachimi} B.,  2019, \mn@doi [\aap]
  {10.1051/0004-6361/201935912}, \href
  {https://ui.adsabs.harvard.edu/abs/2019A&A...631A.160H} {631, A160}

\bibitem[\protect\citeauthoryear{{Harnois-D{\'e}raps}, {Martinet}, {Castro},
  {Dolag}, {Giblin}, {Heymans}, {Hildebrandt}  \& {Xia}}{{Harnois-D{\'e}raps}
  et~al.}{2020}]{Harnois2020}
{Harnois-D{\'e}raps} J.,  {Martinet} N.,  {Castro} T.,  {Dolag} K.,  {Giblin}
  B.,  {Heymans} C.,  {Hildebrandt} H.,   {Xia} Q.,  2020, arXiv e-prints,
  \href {https://ui.adsabs.harvard.edu/abs/2020arXiv201202777H} {p.
  arXiv:2012.02777}

\bibitem[\protect\citeauthoryear{{Harnois-D{\'e}raps}, {Martinet}  \&
  {Reischke}}{{Harnois-D{\'e}raps} et~al.}{2021}]{harnois2021}
{Harnois-D{\'e}raps} J.,  {Martinet} N.,   {Reischke} R.,  2021, arXiv
  e-prints, \href {https://ui.adsabs.harvard.edu/abs/2021arXiv210708041H} {p.
  arXiv:2107.08041}

\bibitem[\protect\citeauthoryear{{Hartlap}, {Simon}  \& {Schneider}}{{Hartlap}
  et~al.}{2007}]{Hartlap2007}
{Hartlap} J.,  {Simon} P.,   {Schneider} P.,  2007, \mn@doi [\aap]
  {10.1051/0004-6361:20066170}, \href
  {https://ui.adsabs.harvard.edu/abs/2007A&A...464..399H} {464, 399}

\bibitem[\protect\citeauthoryear{{Heydenreich}, {Br{\"u}ck}  \&
  {Harnois-D{\'e}raps}}{{Heydenreich} et~al.}{2020}]{Heydenreich2020}
{Heydenreich} S.,  {Br{\"u}ck} B.,   {Harnois-D{\'e}raps} J.,  2020, arXiv
  e-prints, \href {https://ui.adsabs.harvard.edu/abs/2020arXiv200713724H} {p.
  arXiv:2007.13724}

\bibitem[\protect\citeauthoryear{Heymans et~al.,}{Heymans
  et~al.}{2012}]{Heymans2012}
Heymans C.,  et~al., 2012, \mn@doi [Mon. Not. Roy. Astron. Soc.]
  {10.1111/j.1365-2966.2012.21952.x}, 427, 146

\bibitem[\protect\citeauthoryear{Hikage et~al.,}{Hikage
  et~al.}{2019}]{Hikage2019}
Hikage C.,  et~al., 2019, \mn@doi [Pub. Astron. Soc. Japan]
  {10.1093/pasj/psz010}, 71

\bibitem[\protect\citeauthoryear{{Hilbert} et~al.,}{{Hilbert}
  et~al.}{2020}]{Hilbert2020}
{Hilbert} S.,  et~al., 2020, \mn@doi [\mnras] {10.1093/mnras/staa281}, \href
  {https://ui.adsabs.harvard.edu/abs/2020MNRAS.493..305H} {493, 305}

\bibitem[\protect\citeauthoryear{{Hildebrandt} et~al.,}{{Hildebrandt}
  et~al.}{2017}]{Hildebrandt2017}
{Hildebrandt} H.,  et~al., 2017, \mn@doi [Mon. Not. Roy. Astron. Soc.]
  {10.1093/mnras/stw2805}, \href
  {http://adsabs.harvard.edu/abs/2017MNRAS.465.1454H} {465, 1454}

\bibitem[\protect\citeauthoryear{Hoekstra et~al.,}{Hoekstra
  et~al.}{2006}]{Hoekstra2006}
Hoekstra H.,  et~al., 2006, \mn@doi [Astrophys. J.] {10.1086/503249}, 647, 116

\bibitem[\protect\citeauthoryear{{Jain} \& {Van Waerbeke}}{{Jain} \& {Van
  Waerbeke}}{2000}]{Jain2000}
{Jain} B.,  {Van Waerbeke} L.,  2000, \mn@doi [Astrophys. J.] {10.1086/312480},
  \href {https://ui.adsabs.harvard.edu/abs/2000ApJ...530L...1J} {530, L1}

\bibitem[\protect\citeauthoryear{{Kacprzak} et~al.,}{{Kacprzak}
  et~al.}{2016}]{Kacprzak2016}
{Kacprzak} T.,  et~al., 2016, \mn@doi [\mnras] {10.1093/mnras/stw2070}, \href
  {https://ui.adsabs.harvard.edu/abs/2016MNRAS.463.3653K} {463, 3653}

\bibitem[\protect\citeauthoryear{{Kaiser}, {Wilson}  \& {Luppino}}{{Kaiser}
  et~al.}{2000}]{Kaiser2000}
{Kaiser} N.,  {Wilson} G.,   {Luppino} G.~A.,  2000, preprint, \href
  {https://ui.adsabs.harvard.edu/\#abs/2000astro.ph..3338K} {} (\mn@eprint
  {arXiv} {astro-ph/0003338})

\bibitem[\protect\citeauthoryear{{Kilbinger}}{{Kilbinger}}{2015}]{Kilbinger2015}
{Kilbinger} M.,  2015, \mn@doi [Rept. Prog. Phys.]
  {10.1088/0034-4885/78/8/086901}, \href
  {https://ui.adsabs.harvard.edu/abs/2015RPPh...78h6901K} {78, 086901}

\bibitem[\protect\citeauthoryear{{Kilbinger} et~al.,}{{Kilbinger}
  et~al.}{2013}]{Kilbinger2013}
{Kilbinger} M.,  et~al., 2013, \mn@doi [Mon. Not. Roy. Astron. Soc.]
  {10.1093/mnras/stt041}, \href
  {https://ui.adsabs.harvard.edu/\#abs/2013MNRAS.430.2200K} {430, 2200}

\bibitem[\protect\citeauthoryear{{Landy} \& {Szalay}}{{Landy} \&
  {Szalay}}{1993}]{Landy1993}
{Landy} S.~D.,  {Szalay} A.~S.,  1993, \mn@doi [Astrophys. J.]
  {10.1086/172900}, \href {http://adsabs.harvard.edu/abs/1993ApJ...412...64L}
  {412, 64}

\bibitem[\protect\citeauthoryear{{Li}, {Liu}, {Matilla}  \& {Coulton}}{{Li}
  et~al.}{2019}]{Li2018}
{Li} Z.,  {Liu} J.,  {Matilla} J. M.~Z.,   {Coulton} W.~R.,  2019, \mn@doi
  [Phys. Rev. D] {10.1103/PhysRevD.99.063527}, \href
  {https://ui.adsabs.harvard.edu/abs/2019PhRvD..99f3527L} {99, 063527}

\bibitem[\protect\citeauthoryear{{Liu} \& {Haiman}}{{Liu} \&
  {Haiman}}{2016}]{J.Liu2016}
{Liu} J.,  {Haiman} Z.,  2016, \mn@doi [Phys. Rev. D]
  {10.1103/PhysRevD.94.043533}, \href
  {https://ui.adsabs.harvard.edu/\#abs/2016PhRvD..94d3533L} {94, 043533}

\bibitem[\protect\citeauthoryear{{Liu} et~al.,}{{Liu} et~al.}{2016}]{X.Liu2016}
{Liu} X.,  et~al., 2016, \mn@doi [Phys. Rev. Lett.]
  {10.1103/PhysRevLett.117.051101}, \href
  {https://ui.adsabs.harvard.edu/\#abs/2016PhRvL.117e1101L} {117, 051101}

\bibitem[\protect\citeauthoryear{{Marian}, {Smith}, {Hilbert}  \&
  {Schneider}}{{Marian} et~al.}{2013}]{Marian2013}
{Marian} L.,  {Smith} R.~E.,  {Hilbert} S.,   {Schneider} P.,  2013, \mn@doi
  [Mon. Not. Roy. Astron. Soc.] {10.1093/mnras/stt552}, \href
  {https://ui.adsabs.harvard.edu/abs/2013MNRAS.432.1338M} {432, 1338}

\bibitem[\protect\citeauthoryear{{Martinet} et~al.,}{{Martinet}
  et~al.}{2018}]{Martinet2018}
{Martinet} N.,  et~al., 2018, \mn@doi [\mnras] {10.1093/mnras/stx2793}, \href
  {https://ui.adsabs.harvard.edu/abs/2018MNRAS.474..712M} {474, 712}

\bibitem[\protect\citeauthoryear{{Martinet}, {Harnois-D{\'e}raps}, {Jullo}  \&
  {Schneider}}{{Martinet} et~al.}{2020}]{Martinet2020}
{Martinet} N.,  {Harnois-D{\'e}raps} J.,  {Jullo} E.,   {Schneider} P.,  2020,
  arXiv e-prints, \href {https://ui.adsabs.harvard.edu/abs/2020arXiv201007376M}
  {p. arXiv:2010.07376}

\bibitem[\protect\citeauthoryear{{Osato}, {Shirasaki}  \& {Yoshida}}{{Osato}
  et~al.}{2015}]{Osato2015}
{Osato} K.,  {Shirasaki} M.,   {Yoshida} N.,  2015, \mn@doi [Astrophys. J.]
  {10.1088/0004-637X/806/2/186}, \href
  {https://ui.adsabs.harvard.edu/abs/2015ApJ...806..186O} {806, 186}

\bibitem[\protect\citeauthoryear{Pedregosa et~al.,}{Pedregosa
  et~al.}{2011}]{Pedregosa2011}
Pedregosa F.,  et~al., 2011, Journal of Machine Learning Research, 12, 2825

\bibitem[\protect\citeauthoryear{Pen, Zhang, van Waerbeke, Mellier, Zhang  \&
  Dubinski}{Pen et~al.}{2003}]{Pen2003}
Pen U.-L.,  Zhang T.,  van Waerbeke L.,  Mellier Y.,  Zhang P.,   Dubinski J.,
  2003, \mn@doi [Astrophys. J.] {10.1086/375734}, 592, 664

\bibitem[\protect\citeauthoryear{{Planck Collaboration} et~al.,}{{Planck
  Collaboration} et~al.}{2018}]{Planck2018}
{Planck Collaboration} et~al., 2018, arXiv e-prints, \href
  {https://ui.adsabs.harvard.edu/\#abs/2018arXiv180706209P} {}

\bibitem[\protect\citeauthoryear{Riess, Casertano, Yuan, Macri  \&
  Scolnic}{Riess et~al.}{2019}]{Riess:2019cxk}
Riess A.~G.,  Casertano S.,  Yuan W.,  Macri L.~M.,   Scolnic D.,  2019,
  \mn@doi [Astrophys. J.] {10.3847/1538-4357/ab1422}, 876, 85

\bibitem[\protect\citeauthoryear{{Schneider}, {van Waerbeke}, {Kilbinger}  \&
  {Mellier}}{{Schneider} et~al.}{2002}]{Schneider2002}
{Schneider} P.,  {van Waerbeke} L.,  {Kilbinger} M.,   {Mellier} Y.,  2002,
  \mn@doi [\aap] {10.1051/0004-6361:20021341}, \href
  {https://ui.adsabs.harvard.edu/\#abs/2002A&A...396....1S} {396, 1}

\bibitem[\protect\citeauthoryear{Schneider, Knox, Habib, Heitmann, Higdon  \&
  Nakhleh}{Schneider et~al.}{2008}]{Schneider2008}
Schneider M.~D.,  Knox L.,  Habib S.,  Heitmann K.,  Higdon D.,   Nakhleh C.,
  2008, \mn@doi [Phys. Rev. D] {10.1103/PhysRevD.78.063529}, 78, 063529

\bibitem[\protect\citeauthoryear{{Semboloni} et~al.,}{{Semboloni}
  et~al.}{2006}]{Semboloni2006}
{Semboloni} E.,  et~al., 2006, \mn@doi [\aap] {10.1051/0004-6361:20054479},
  \href {https://ui.adsabs.harvard.edu/\#abs/2006A&A...452...51S} {452, 51}

\bibitem[\protect\citeauthoryear{Troxel et~al.,}{Troxel
  et~al.}{2018}]{Troxel2018}
Troxel M.~A.,  et~al., 2018, \mn@doi [Phys. Rev. D]
  {10.1103/PhysRevD.98.043528}, 98, 043528

\bibitem[\protect\citeauthoryear{{Van Waerbeke} et~al.,}{{Van Waerbeke}
  et~al.}{2000}]{VanWaerbeke2000}
{Van Waerbeke} L.,  et~al., 2000, \aap, \href
  {http://adsabs.harvard.edu/abs/2000A\%26A...358...30V} {358, 30}

\bibitem[\protect\citeauthoryear{{Verde}, {Treu}  \& {Riess}}{{Verde}
  et~al.}{2019}]{Verde2019}
{Verde} L.,  {Treu} T.,   {Riess} A.~G.,  2019, \mn@doi [Nature Astronomy]
  {10.1038/s41550-019-0902-0}, \href
  {https://ui.adsabs.harvard.edu/abs/2019NatAs...3..891V} {3, 891}

\bibitem[\protect\citeauthoryear{{Wei}, {Li}, {Kang}, {Liu}, {Fan}, {Yuan}  \&
  {Pan}}{{Wei} et~al.}{2018}]{Wei2018}
{Wei} C.,  {Li} G.,  {Kang} X.,  {Liu} X.,  {Fan} Z.,  {Yuan} S.,   {Pan} C.,
  2018, \mn@doi [Mon. Not. Roy. Astron. Soc.] {10.1093/mnras/sty1268}, \href
  {https://ui.adsabs.harvard.edu/\#abs/2018MNRAS.478.2987W} {478, 2987}

\bibitem[\protect\citeauthoryear{{Weiss}, {Schneider}, {Sgier}, {Kacprzak},
  {Amara}  \& {Refregier}}{{Weiss} et~al.}{2019}]{Weiss2019}
{Weiss} A.~J.,  {Schneider} A.,  {Sgier} R.,  {Kacprzak} T.,  {Amara} A.,
  {Refregier} A.,  2019, \mn@doi [J. Cosmo. Astropart. Phys.]
  {10.1088/1475-7516/2019/10/011}, \href
  {https://ui.adsabs.harvard.edu/abs/2019JCAP...10..011W} {2019, 011}

\bibitem[\protect\citeauthoryear{{White}}{{White}}{1979}]{White1979}
{White} S.~D.~M.,  1979, \mn@doi [Mon. Not. Roy. Astron. Soc.]
  {10.1093/mnras/186.2.145}, \href
  {https://ui.adsabs.harvard.edu/abs/1979MNRAS.186..145W/abstract} {186, 145}

\bibitem[\protect\citeauthoryear{{Wittman}, {Tyson}, {Kirkman}, {Dell'Antonio}
  \& {Bernstein}}{{Wittman} et~al.}{2000}]{Wittman2000}
{Wittman} D.~M.,  {Tyson} J.~A.,  {Kirkman} D.,  {Dell'Antonio} I.,
  {Bernstein} G.,  2000, \mn@doi [\nat] {10.1038/35012001}, \href
  {http://adsabs.harvard.edu/abs/2000Natur.405..143W} {405, 143}

\bibitem[\protect\citeauthoryear{{Yang}, {Kratochvil}, {Wang}, {Lim}, {Haiman}
  \& {May}}{{Yang} et~al.}{2011}]{Yang2011}
{Yang} X.,  {Kratochvil} J.~M.,  {Wang} S.,  {Lim} E.~A.,  {Haiman} Z.,   {May}
  M.,  2011, \mn@doi [Phys. Rev. D] {10.1103/PhysRevD.84.043529}, \href
  {https://ui.adsabs.harvard.edu/\#abs/2011PhRvD..84d3529Y} {84, 043529}

\makeatother
\end{thebibliography}

%%%%%%%%%%%%%%%%%%%%%%%%%%%%%%%%%%%%%%%%%%%%%%%%%%

%%%%%%%%%%%%%%%%% APPENDICES %%%%%%%%%%%%%%%%%%%%%

\appendix

\section{Emulator accuracy}
\label{app:accuracy}

In this section we present the accuracy of the peak abundance and peak 2PCF emulator used for our cosmological forecasts. 

In order to test the accuracy of the emulator, we employ a cross validation test, which is outlined as follows. First, one node from the training data (simulated data) is removed, and the emulator is then trained with the remaining 25 nodes, for a given statistics. The emulator prediction for the removed node is then calculated, and this result is compared to the simulated version, by taking the difference between the two and dividing it by the standard error measured in the simulated data for that node. The above steps are repeated 25 more times, removing a different node from the training data at each iteration. This results in measurements of the emulator accuracy at each node, which is an upper limit, since the accuracy increases as more training data is used, and the cross validation measurements uses training data with one less node than the training data used in the main analysis.

\begin{figure}
    \centering
    \includegraphics[width=\columnwidth]{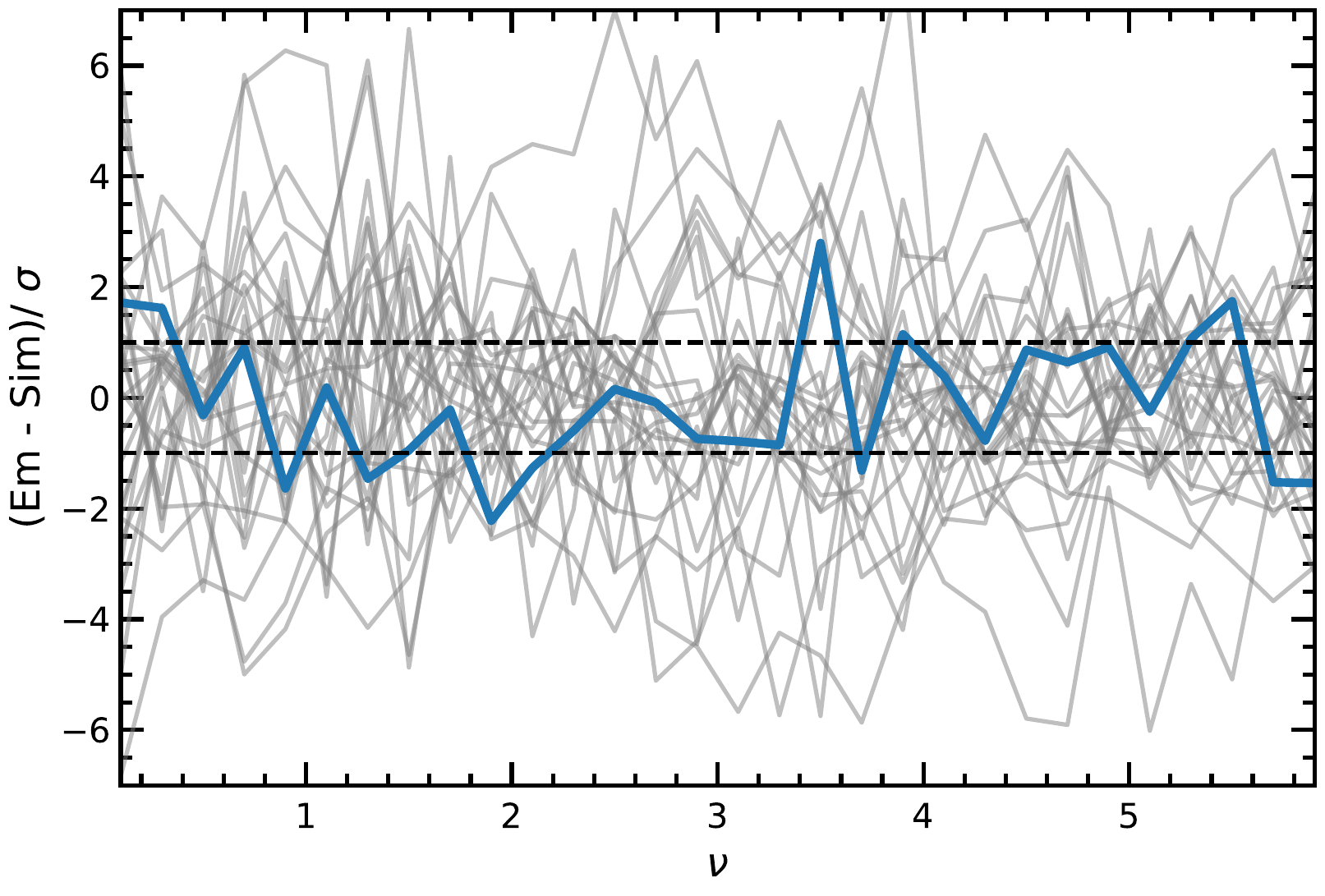}
    \caption{(Colour Online) The cross validation of the WL peak abundance emulator accuracy. One node is removed from the training set, and the difference between the emulation (Em) and simulation (Sim) predictions of the removed node are compared relative to the simulation standard error ($\sigma$). This process is repeated for each of the 26 nodes, giving an upper limit on the emulator accuracy. The iteration where the node for the fiducial cosmology is removed is shown by the blue line. Dashed lines are added at the $1\sigma$ level to help guide the reader. }
    \label{fig:PA emulator accuracy}
\end{figure}

\begin{figure}
    \centering
    \includegraphics[width=\columnwidth]{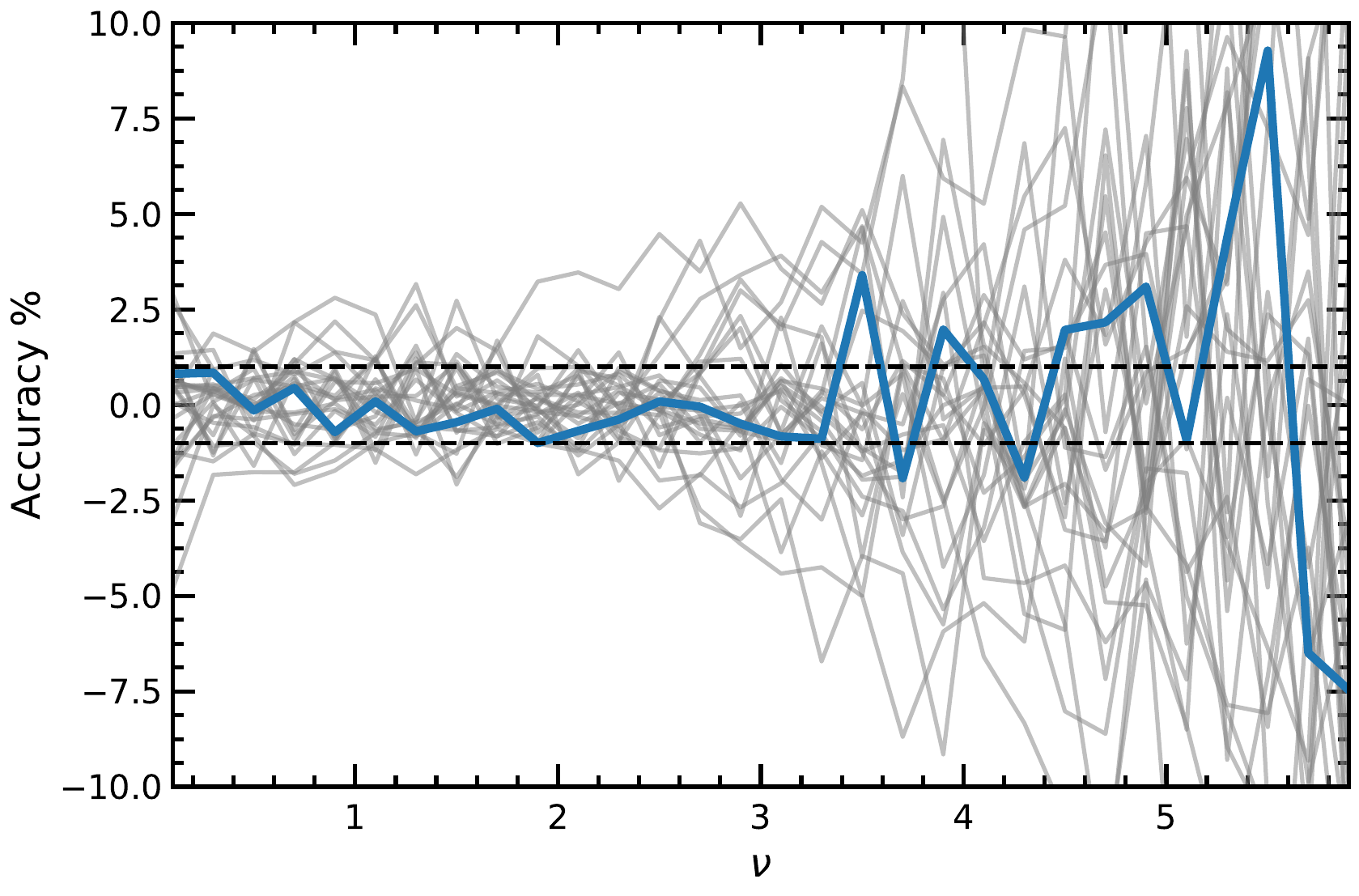}
    \caption{(Colour Online) The same as Fig. \ref{fig:PA emulator accuracy}, but showing the percentage accuracy relative to the simulated predictions. The dashed lines enclose the $1\%$ region.}
    \label{fig:PA emulator accuracy perc}
\end{figure}

Fig.~\ref{fig:PA emulator accuracy} shows the cross validation test for WL peak abundance. The cross validation for the fiducial cosmology is shown in blue, and the remaining nodes are shown in grey. The fiducial cosmology is the node of most interest, as all posterior contours presented in this work reside close to this region. The dashed lines delineate the region where the accuracy of the emulator is within the standard error of the simulated data. The blue curve shows that, for the fiducial cosmology, the emulator is accurate to within $1\sigma$, as roughly $68\%$ of the bins are within the $1\sigma$ region. The grey curves show that the accuracy is lower for the other nodes, and we find that the accuracy decreases as we approach the edges of the \cosmoslics{} parameter space. This is to be expected, as the emulator has less data to train from for these regions. Fig.~\ref{fig:PA emulator accuracy perc} is similar to Fig.~\ref{fig:PA emulator accuracy}, but shows the percentage accuracy of the emulator for the cross validation test. We can observe that the accuracy is within roughly $1\%$ for $\nu\lesssim3$, increasing to up to $4\%$ at $\nu\gtrsim5$ due to the more noisy measurement for the high-$\nu$ peaks.

\begin{figure*}
    \centering
    \includegraphics[width=2\columnwidth]{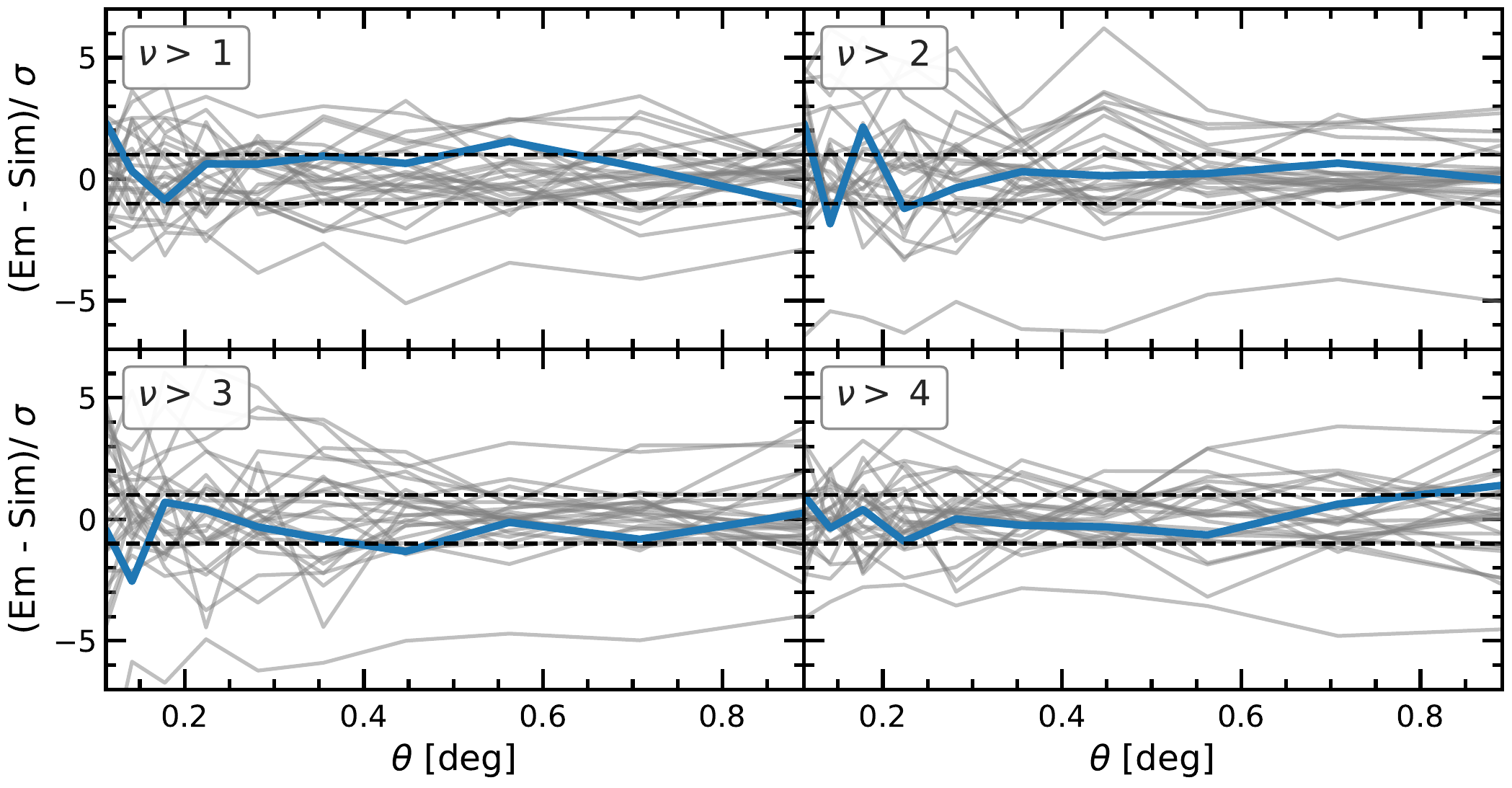}
    \caption{The same as Fig.~\ref{fig:PA emulator accuracy} but for the WL peak 2PCFs. The four panels correspond to the WL peak 2PCFs of WL peaks with heights $nu > 1$ (top left), $\nu > 2$ (top right), $\nu > 3$ (bottom left) and $\nu > 4$ (bottom right).  }
    \label{fig:P2PCF emulator accuracy}
\end{figure*}

\begin{figure*}
    \centering
    \includegraphics[width=2\columnwidth]{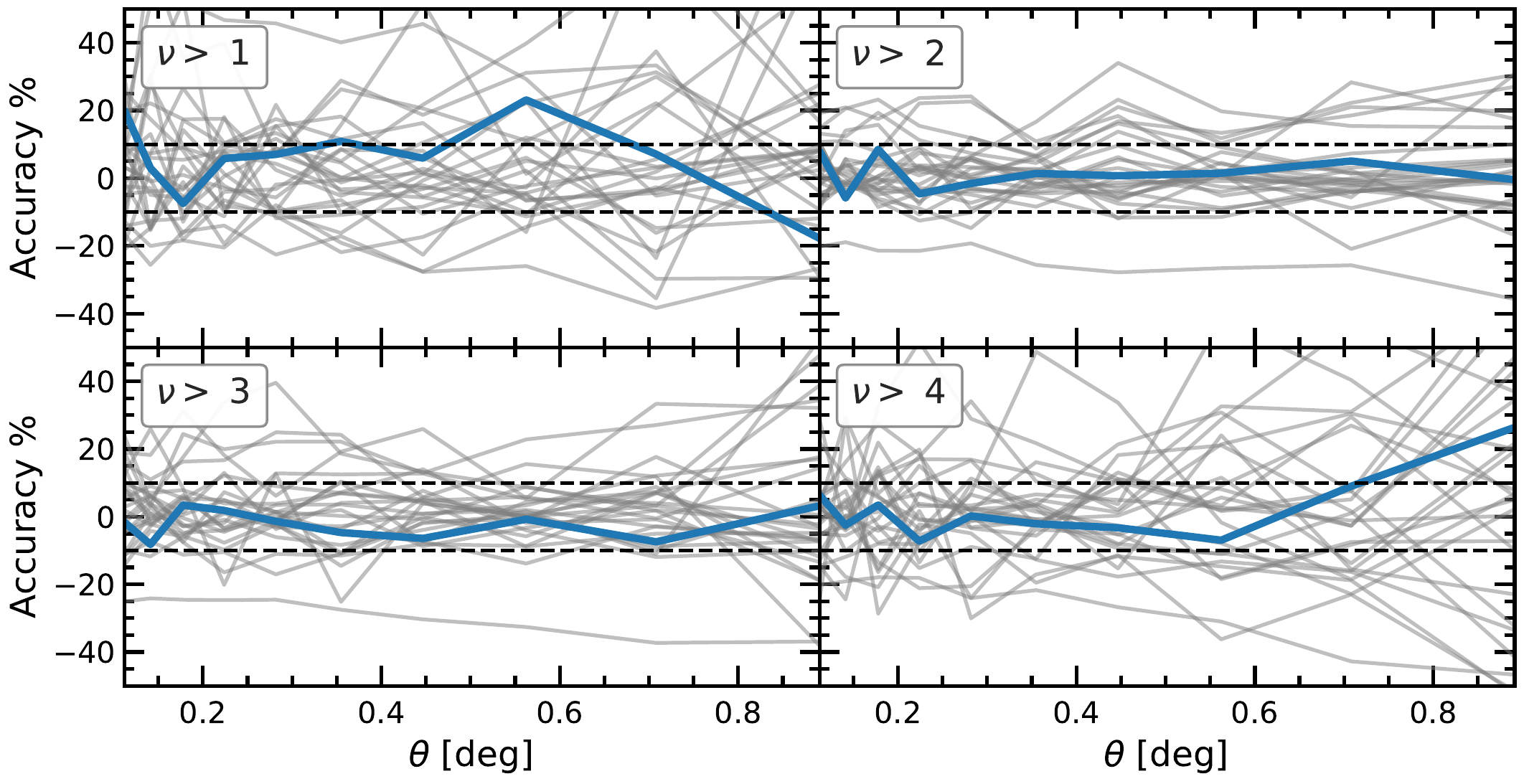}
    \caption{The same as Fig.~\ref{fig:P2PCF emulator accuracy} but for the percentage accuracy relative to the simulated predictions. The dashed lines enclose the $10\%$ region. }
    \label{fig:P2PCF emulator accuracy perc}
\end{figure*}

Fig.~\ref{fig:P2PCF emulator accuracy} is the same as Fig. \ref{fig:PA emulator accuracy}, but shows the cross validation test for the WL peak 2PCF, for peak catalogues with heights $\nu > 1$ (top left), $\nu > 2$ (top right), $\nu > 3$ (bottom left) and $\nu > 4$ (bottom right). The figure shows that, similar to the WL peak abundance, the emulator is accurate to within $1\sigma$ at the fiducial cosmology for the peak 2PCF for all four catalogues. Fig.~\ref{fig:P2PCF emulator accuracy perc} is the same as Fig.~\ref{fig:P2PCF emulator accuracy}, but shows the percentage accuracy of the peak 2PCF emulator applied in the cross-validation test. For the $\nu > 1$ catalogue, the accuracy is mostly within $10\%$, with a few bin at $20\%$. The accuracy is within $10\%$ for the $\nu > 2$ and $3$ peak catalogues, and for $\nu > 4$ the accuracy is within $10\%$ except for the final bin.

\section{Covariance}\label{app:cov}

\begin{figure*}
    \centering
    \includegraphics[width=2\columnwidth]{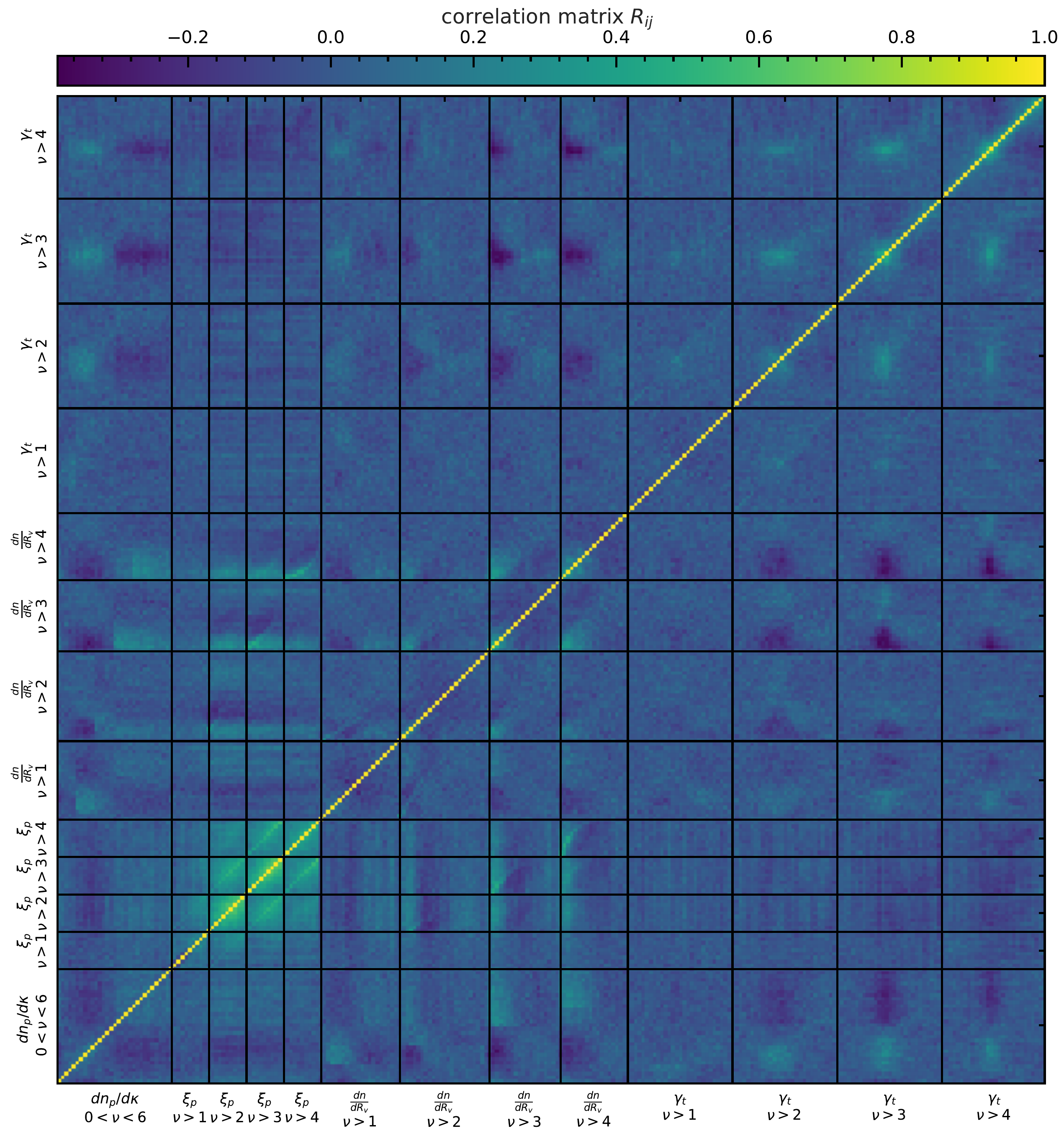}
    \caption{(Colour Online) The correlation matrix for all probes studied in this work, which are as follows (from left to right): peak abundance ($dn/d\kappa$), peak 2PCF ($\xi_p$), void abundance ($dn/dR_v$) and void tangential shear profiles ($\gamma_t$).}
    \label{fig:covariance}
\end{figure*}

\begin{figure*}
    \centering
    \includegraphics[width=2\columnwidth]{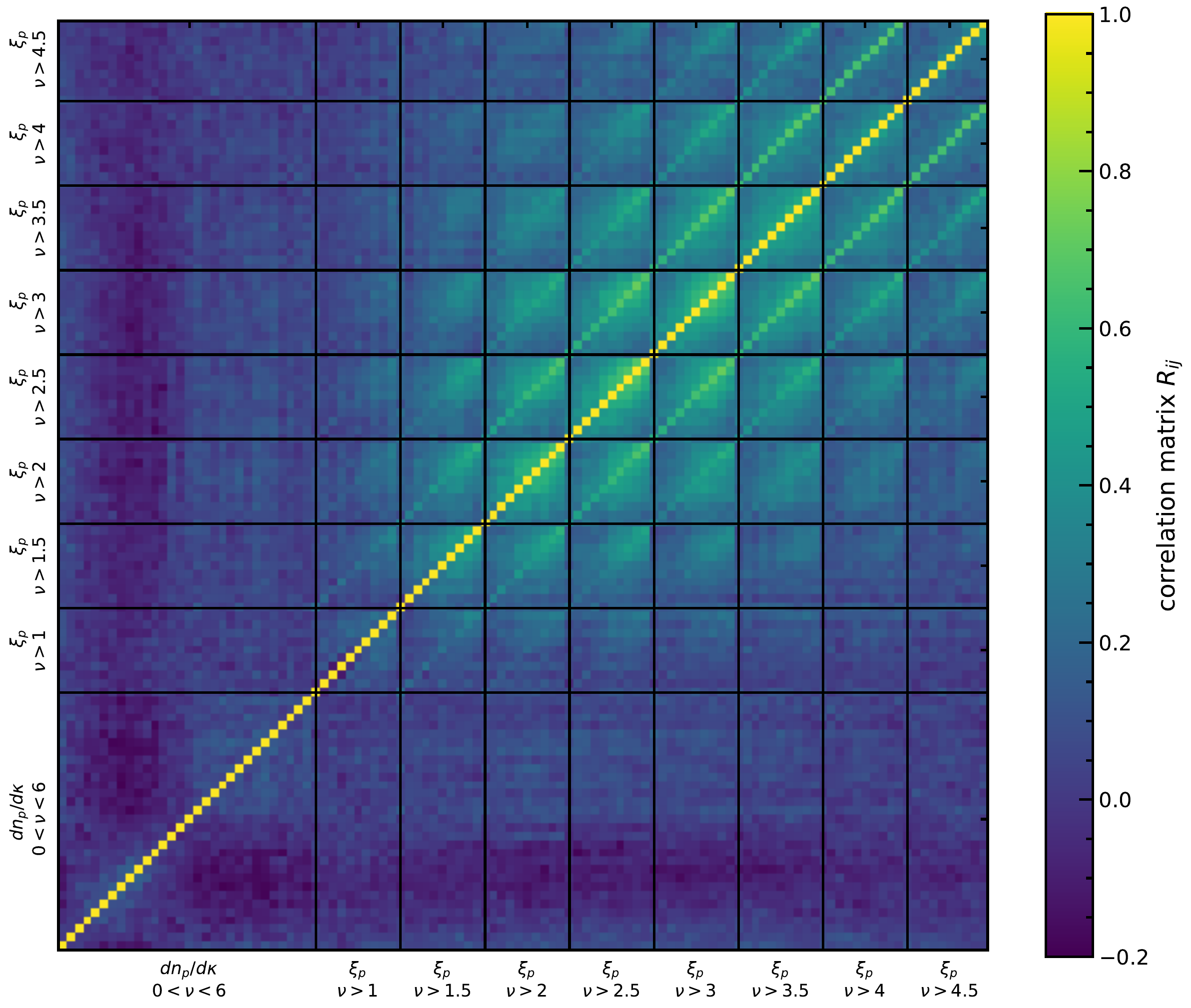}
    \caption{(Colour Online) The correlation matrix for the combination of the peak abundance ($dn/d\kappa$) and the peak 2PCF ($\xi_p$, for eight catalogues with $\nu> 1.0,1.5,...,4.5$). }
    \label{fig:covariance peaks}
\end{figure*}

As shown by Eq.~\eqref{eq:log likelihood} we require the (inverted) covariance matrix of the data vector in order to produce our forecasts. Within the matrix, the diagonal elements correspond to the variance of each of the data vector bins and the off-diagonal elements give the covariance between all possible pairs of bins. 
When multiple probes are combined into a single data vector, any correlated or duplicate information between the probes is accounted for by the cross covariance within the matrix.

In Fig.~\ref{fig:covariance} we present the total correlation matrix for the all of the probes studied in this work. This corresponds to the matrix that is used to produce the green likelihood contour in Fig. \ref{fig:peak and void contours comparison}. We present the correlation matrix instead of the covariance matrix, as it allows for easier visual interpretation, which is expressed in terms of the covariance matrix as follows
\begin{equation}
    R_{ij} = \frac{{\rm cov}(i,j)}{\sigma_i\sigma_j} \, , 
\end{equation}
where $R$ is the correlation matrix, ${\rm cov}$ is the covariance matrix and $\sigma$ is the standard deviation for a given bin.

Starting from the bottom left of the figure, the diagonal tiles enclosed by the black lines show the correlation for the following statistics (which are labelled with the range of peak heights used in their identification): peak abundance ($0<\nu<6$), peak 2PCF (four catalogues with thresholds $\nu>1,2,3,4$), and WL void abundance and WL void tangential shear profiles where the voids are identified using the same four peak catalogues. The remaining off-diagonal terms show the cross-covariances between all possible combinations of the probes. 

For the peak abundance, the figure shows that the low amplitude peaks are somewhat correlated with other low amplitude peaks, and a similar behaviour is present for the high amplitude peaks, as shown by the green regions in the bottom left and top right of the peak abundance correlation tile. There also appears to be a small amount of anti-correlation between low and high amplitude peaks, as shown by the dark regions in the top left and bottom right of the tile. 

For the diagonal peak 2PCF tiles, each bin in the 2PCF appears to be correlated with all of the other bins. For the off-diagonal tiles between the different peak 2PCFs, there is also a high amount of correlation, which is again expected, as the main difference between the 2PCFs is simply a change in amplitude, and all catalogues have some fraction of their tracer population in common. 

For the tiles representing the correlation between the peak 2PCFs and the WL void abundances, we see some correlation between the peak 2PCFs and the small radii WL voids (especially for high $\nu$ thresholds). This is also to be expected since the WL voids are identified from a Delaunay triangulation of the peaks, which will be sensitive to the peak clustering. It is interesting to see that this correlation drops off as the void size increases, which may indicate that higher-order clustering such as the three-point correlation function of WL peaks dictates the abundance of large voids.

In Fig. \ref{fig:covariance peaks} we show the correlation matrix for the peak abundance combined with the eight peak 2PCFs with $\nu > 1.0,1.5,...,4.5$. The figure shows that, for the peak 2PCF, adjacent catalogues (similar $\nu$ thresholds) are highly correlated. This is to be expected as the tracer populations are very similar for adjacent catalogues. The correlation reduces significantly as the difference between the $\nu$ thresholds increases, which is again expected as this is where the tracer populations will differ the most. The low correlation between the peak 2PCFs with very different $\nu$ thresholds is a strong contribution to the improved constraining power from the combination of multiple peak 2PCFs, alongside any complementary parameter degeneracy directions.

%%%%%%%%%%%%%%%%%%%%%%%%%%%%%%%%%%%%%%%%%%%%%%%%%%

% Don't change these lines
\bsp	% typesetting comment
\label{lastpage}
\end{document}